\documentclass[journal]{IEEEtran} %%
\usepackage{mathrsfs}
\usepackage{amssymb}
\usepackage[mathscr]{eucal}
\usepackage{makecell}
\usepackage{cite}
\usepackage{graphicx}
\usepackage{psfrag}
\usepackage{subfigure}
\usepackage{url}
\usepackage{stfloats}
\usepackage{amsmath}
\usepackage{float}
\usepackage{array}
\usepackage{amsthm}
\usepackage{booktabs}  %
\usepackage{algorithm} %format of the algorithm
\usepackage{algorithmic} %format of the algorithm
\usepackage{color}
\usepackage{cases}
\usepackage{bm}
\usepackage{changepage}
\usepackage{amsmath,amsfonts,amssymb,amsbsy,paralist,ifthen}
\allowdisplaybreaks[4]

\begin{document}

% paper title
\title{STAR-RIS Enabled ISAC Systems:
Joint Rate Splitting and Beamforming Optimization
}

\author{Yuan~Liu,
~Ruichen~Zhang,~
Ruihong~Jiang,
~Yongdong~Zhu,
~Huimin~ Hu,~\\
Qiang~Ni,~\IEEEmembership{Senior Member,~IEEE},
~Zesong Fei,~\IEEEmembership{Senior Member,~IEEE},
~Dusit Niyato,~\IEEEmembership{Fellow,~IEEE}

\thanks{
%This work was supported in part by the National Natural Science Foundation of China (NSFC) under Grant No. 62301077 and No. 62401513,
%in part by the National Key R\&D Program of China under Grant No. 2021YFB2900200,
%in part by the Fundamental Research Funds for Universities under grant No. 2023RC17.
%% and by the China Postdoctoral Science Foundation under Grant No.2023M733274.
%\emph{(Corresponding author: Yongdong Zhu.)}

Yuan Liu and Yongdong Zhu are with the Research Center for Space Computing System, Zhejiang Laboratory, Hangzhou, 311100, China (e-mail: liuyuan@bjtu.edu.cn, zhuyongdong@hotmail.com).

Ruichen Zhang and
Dusit Niyato are with the College of Computing and Data Science, Nanyang Technological University, Singapore 639798
(email: ruichen.zhang@ntu.edu.sg, dniyato@ntu.edu.sg).

Ruihong Jiang is with the State Key Laboratory of Networking and Switching Technology, Beijing University of Posts and Telecommunications, Beijing 100876, China
(e-mail: rhjiang@bupt.edu.cn).

Huimin Hu is with the School of Communications and Information Engineering and the School of Artificial Intelligence, Xi'an University of Posts and Telecommunications, Xi'an 710121, China
(e-mail:huiminhu@xupt.edu.cn).

Qiang Ni is with the School of Computing and Communications, Lancaster University, Lancaster, LA1 4WA, United Kingdom (e-mail: q.ni@lancaster.ac.uk).

Zesong Fei is with the School of Information and Electronics, Beijing Institute of Technology, Beijing 100081, China (E-mail: feizesong@bit.edu.cn).

}
}

\maketitle

\begin{abstract}
This paper delves into
an integrated sensing and communication (ISAC) system bolstered by a simultaneously transmitting and reflecting reconfigurable intelligent surface (STAR-RIS).
Within this system, a base station (BS) is equipped with communication and radar capabilities, enabling it to communicate with ground terminals (GTs) and concurrently probe for echo signals from a target of interest.
Moreover,
to manage interference and improve communication quality,
the rate splitting multiple access (RSMA) scheme
is incorporated into the system.
The signal-to-interference-plus-noise ratio (SINR)
of the received sensing echo signals is a measure of sensing performance.
We formulate a joint optimization problem of common rates, transmit beamforming at the BS, and passive beamforming vectors of the STAR-RIS. The objective is to maximize sensing SINR while guaranteeing the communication rate requirements for each GT.
We present an iterative algorithm to
address the non-convex problem by invoking
Dinkelbach's transform,
semidefinite relaxation (SDR),
majorization-minimization,
and sequential rank-one constraint relaxation (SROCR) theories.
Simulation results manifest that
the performance of the studied ISAC network enhanced by the STAR-RIS and RSMA surpasses other benchmarks considerably.
The results evidently indicate
the superior performance improvement of the ISAC system
with the proposed RSMA-based transmission strategy design
and the dynamic optimization of both transmission and
reflection beamforming at STAR-RIS.
%In addition,
%it confirms that
%the sensing performance gain obtained by employing an independent communication waveform design is comparable to that obtained by a
%joint communication and sensing waveform design.

\end{abstract}


\section{Introduction}
\subsection{Background}
Recently,
a large number of new intelligent applications have emerged,
such as autonomous vehicles, smart industry, and cellular networks
for 5G and beyond,
which have increasingly stringent
communication requirements for high bandwidth and high transmission capacity, as well as perception requirements for high-precision and high-resolution \cite{intro_mag_v1}
\cite{intro_Iot}.
Meanwhile,
because of the restricted accessibility of spectrum resources
coupled with the communication performance progressively nearing its theoretical limit,
the research into integrated sensing and communication (ISAC) systems has consistently gained momentum,
attracting strong attention
from both academic and industrial sectors
\cite{intro_mag_v2,intro_mag_v3,intro_mag_v4}.
Precisely,
the key idea of ISAC is to integrate both communication and sensing functionalities over shared
time-frequency-power-hardware resources in one single system \cite{intro_mag_v5}.
By leveraging a unified signal processing framework, spectrum, and hardware platform,
ISAC technology has the potential to boost spectral and energy efficiencies,
thereby tackling spectrum congestion and resource wastage issues,
while simultaneously reducing hardware and signalling costs \cite{intro_mag_v6}.
Thus,
the ISAC technology is meaningful for
supporting diverse applications to access wireless networks and meet their high-quality wireless communications and high-accuracy sensing requirements \cite{intro_mag_v7}.

Furthermore,
as the number of ground terminals (GTs), such as autonomous vehicles and intelligent robots increases,
inter-user interference emerges as a substantial factor that inhibits communication performance.
Fortunately,
rate-splitting multiple access (RSMA) has been proposed,
which is widely recognized as a promising manner for achieving robust interference management and communication enhancement \cite{intro_RSMA_v1}.
% It also has garnered extensive research attention across ISAC wireless networks.
Using the RSMA scheme at the transmitter,
the information streams are selectively encoded into a shared common stream and individual private streams by means of linear precoded rate-splitting \cite{intro_RSMA_proposed}.
Particularly,
the common stream should be decoded by
all receivers,
while the private streams are required to be encoded independently and decoded by the corresponding receivers with successive interference cancellation (SIC) \cite{intro_RSMA_introduction}.
By this way,
the RSMA scheme can
alleviate the tensions arising from the scarcity of wireless resources
and multi-user communication requirements,
thereby improving the performance of communication systems including ISAC
\cite{intro_RSMA_v2}.

On the other hand,
sensing performance is also an important indicator for ISAC systems,
which may be restricted by severe path loss fading.
In this regard,
reconfigurable intelligent surface (RIS) can construct additional transmission links and enhance the signal strength of desired directions
by simultaneously manipulating the amplitudes and phases of reflective elements \cite{intro_RIS_v0_ly}.
Thus,
RIS can assist in signal enhancement for sensing direction
and provide new degrees of freedom (DoF) for ISAC system designs \cite{intro_RIS_v1}.
% Therefore, by simultaneously manipulating the amplitudes and phases coefficients of the STAR-RIS elements, the impinging signals can be reflected and transmitted toward the desired spatial directions, potentially mitigating some of the severe path loss fading and providing new degrees of freedom (DoF) for ISAC system designs \cite{intro_star_RIS_v2}.
However,
since the general reflecting-only and transmitting-only RISs can only provide half-space coverage of $180^{\circ}$,
GTs distributed on one side of the RIS only be isolated
due to the geographical restriction.
So far,
relying on the superiority of
the simultaneous transmitting and reflecting reconfigurable intelligent surface (STAR-RIS)
for
providing full-space signal coverage of $360^{\circ}$,
it has been devolved into different networks  \cite{intro_star_RIS_v1}.
In particular,
the STAR-RIS possesses the ability to bifurcate the incident signal,
simultaneously directing one segment as transmitted signals and another as reflected signals
thereby providing services to users on both sides of STAR-RIS \cite{intro_star_RIS_v2}.
Therefore, compared with traditional RISs,
STAR-RIS has superior versatility in network deployment due to its comprehensive spatial coverage,
and can also provide enhanced signal propagation towards sensing targets and GTs with a higher DoF \cite{intro_star_RIS_v3}.

\subsection{Related Work}
Recently,
considerable efforts have been devoted to developing ISAC systems empowered by RIS and RSMA.
In \cite{intro_RIS_ISAC_ref1},
a RIS-assisted MIMO ISAC system was taken into account,
where the waveform and passive beamforming
were collaboratively designed
with the goal of elevating the SINR of radar,
while mitigating multi-user interference during communication.
In \cite{intro_RIS_ISAC_ref2},
the RIS-aided ISAC system was studied,
focusing on the investigation of robust beamforming and RIS phase shifts design with the aim of maximizing radar mutual information.
However,
in \cite{intro_RIS_ISAC_ref1,intro_RIS_ISAC_ref2},
only the traditional RISs were considered
instead of STAR-IRS,
the achievable system performance gain is limited.
At present,
STAR-RIS with full spatial coverage has been integrated into ISAC systems in many studies.
In \cite{intro_star_RIS_ISAC_ref1_sdma},
the communication rate and sensing power
were concurrently maximized for the STAR-RIS assisted ISAC system.
% where the transmit beamforming and phase shift of STAR-RIS were both optimized.
In \cite{intro_star_RIS_ISAC_ref2_sdma},
in pursuit of attaining the optimal sensing SINR of ISAC network,
they simultaneously refined the transmit beamforming at the base station (BS) and meticulously adjusted the transmission and reflection beamforming configurations of the STAR-RIS.
In \cite{intro_star_RIS_ISAC_ref3_sdma},
the STAR-RIS was utilized to assist communication capability, while the passive RIS was leveraged to improve sensing functionality,
where the weighted sum-rate of communication users were maximized
by jointly optimizing the beamforming at ISAC BS, and phase shift vector
of STAR-RIS and passive RIS.
However, the transmission scheme based on space division multiple access (SDMA) was adopted in \cite{intro_star_RIS_ISAC_ref1_sdma,intro_star_RIS_ISAC_ref2_sdma, intro_star_RIS_ISAC_ref3_sdma},
which is difficult to provide effective interference suppression
when the number of users increases.

%\textcolor[rgb]{0.00,0.07,1.00}{Owing to the NOMA-based scheme's proficiency in enhancing spectrum efficiency and ensuring user fairness, it has been extensively implemented in ISAC systems.
%In \cite{intro_star_RIS_ISAC_ref_NOMA1},
%the STAR-RIS aided ISAC system with NOMA transmission scheme was studied,
%where a matching error minimization problem was solved to
%find the desirable sensing beam-pattern
%by refining the active and passive beamforming, power and time allocation ratios.
%In \cite{intro_star_RIS_ISAC_ref_NOMA2},
%the ISAC system incorporating the NOMA transmission scheme and aided by STAR-RIS was examined,
%where a problem emphasized on maximizing the minimum beampattern gain was investigated.
%In \cite{intro_star_RIS_ISAC_ref_NOMA3},
%the ISAC system bolstered by the integration
%of STAR-RIS and NOMA-based scheme was scrutinized,
%where the fairness between communication performance and sensing performance was elevated
%by concurrently refining the BS's transmit beamforming and modifying the coefficient matrices at STAR-RIS.}
Meanwhile,
driven by RSMA's capability to mitigate interference,
certain studies have delved into integrating RSMA into the ISAC system.
In \cite{intro_RSMA_ISAC_ref1},
the cooperative ISAC system with RSMA
transmission scheme was investigated,
where the performance region built on the system sum rate and
the boundary limit of positioning error for radar target were characterized.
In \cite{intro_RSMA_ISAC_ref2},
the uplink RSMA enabled ISAC system was studied,
where the transmitted and received beamforming were jointly optimized to realize the optimal sensing SINR,
at the same time guaranteeing the fulfillment of users' rate demands.
In \cite{intro_RSMA_ISAC_ref3},
the dual-functional radar-communication system assisted by RSMA approach was contemplated,
where the message splitting, precoders for communication streams, and radar sequences were collaboratively devised to optimize the weighted sum rate and minimize the mean square error in radar beampattern approximation.
In \cite{intro_RSMA_ISAC_ref4},
a RSMA-powered ISAC system was researched,
emphasizing the minimization of the Cram¨¦r-Rao lower bound (CRLB) with respect to the sensing response matrix,
which was achieved through the design of RSMA structure and associated parameters.
% , all the while maintaining strict adherence to data rate necessities and transmit power limitations.
In \cite{intro_RSMA_ISAC_RIS_ref1},
the RIS-aided ISAC system incorporating the RSMA approach was analyzed,
where the sensing SNR was elevated through
meticulous design of rate splitting coupled with precise adjustments of beamforming at BS and STAR-RIS, respectively.
Nevertheless, in \cite{intro_RSMA_ISAC_ref1,intro_RSMA_ISAC_ref2,intro_RSMA_ISAC_ref3,
intro_RSMA_ISAC_ref4,
intro_RSMA_ISAC_RIS_ref1},
the STAR-RISs with both
reflecting and transmission functionalities was not involved.

\subsection{Motivation and Contributions}
In this paper,
the BS assisted by STAR-RIS
provides communication services for GTs based on the RSMA transmission scheme,
and concurrently performs target sensing
through the beamforming design.
Specifically,
the STAR-RIS employs the energy splitting (ES) mode to partition the incident signal,
directing a portion into the transmission space for sensing the target and another portion into the reflection space for communicating with GTs.
The contributions are outlined as follows.
% Particularly, the STAR-RIS divides the whole space into two portions, namely, the transmission space with all communication users and the reflection space with the sensing target. Moreover, compared with the MS and TS protocols, the ES enjoys both
% higher performance gain and lower operational complexity [16], thus the STAR-RIS adopts the ES instead of others.
\begin{itemize}
\item
Regarding the STAR-RIS-enhanced ISAC system incorporating the RSMA scheme,
the SINR of the received sensing echo signals
is treated as a measure of sensing performance.
We formulate an optimization problem aimed at maximizing the sensing SINR
by jointly optimizing the rate splitting for the common stream, the transmit beamforming
at BS, and the passive beamforming at STAR-RIS, respectively.
To the best of our knowledge,
this is the first work integrating RSMA and STAR-RIS in an ISAC system.

\item
The considered optimized problem involves signal coordination,
interference management, and amplitude adjustment,
and the optimization variables are coupled together.
This leads to the formulated problem being non-convex and difficult to handle.
To address this, firstly,
the primary problem is decomposed into two sub-problems.
Secondly,
the variable substitution, semidefinite relaxation (SDR),
first-order Taylor expansion,
dinkelbach's transform,
and the sequential rank-one constraint relaxation (SROCR) are introduced to deal with the first sub-problem.
Through this approach,
the optimized transmit beamforming of the BS and common-stream rate allocation are obtained.
Thirdly,
the SDR and SROCR methods are also used to tackle the second sub-problem,
and the optimized transmission and reflection beamforming matrices of STAR-RIS
are obtained.
Ultimately,
by iteratively alternating and solving two sub-problems,
we can obtain the solution of the original problem.

\item
Simulation results demonstrate the efficacy of the proposed algorithm in addressing the non-convex problem.
It also reveals that the
transmission and reflection
beamforming design of STAR-RIS and RSMA-based scheme play an important role in enhancing the performance of the ISAC system.
Besides,
we discover that in the examined ISAC network encompassing a single target,
the sensing SINR in the case of transmitting communication signals only is the same as that in the case of transmitting both communication and sensing signals simultaneously.
That's to say,
from the perspective of sensing SINR,
dedicated sensing waveforms are not always necessary.
This finding significantly simplifies the implementation complexity of the network under investigation.

\end{itemize}

The remainder of this paper is structured as follows.
Section \ref{sys model} puts
forward the STAR-RIS enabled ISAC system model.
Section \ref{problem_formulation} formulates an optimization problem for maximizing
the sensing SINR,
and proposes an iterative algorithm for solving the formulated problem.
Section \ref{simulation results}
analyzes the simulation results.
Section \ref{conclusion} concludes this work.

% However,
% wireless communication and radar sensing technology showed a separate and independent development model before.
% This independent system design has problems such as high hardware cost, low spectrum utilization rate, resource waste, limited performance improvement, and cannot meet the multi-dimensional performance requirements of the above intelligent applications.
% To this end,
% the reasarch of integrated sensing and communication (ISAC) has emerged and attracted widespread attention.
% To be specific, the key idea of ISAC is to integrate both communication and sensing functionalities over shared single time-frequency-power-hardware resource,
% so as to improve the system spectral efficiency
% and hardware utilization.
% (introduction of ISAC is too much )

\section{system model}\label{sys model}
\subsection{Network Model}\label{net model}
\begin{figure}[t]
\centering
\includegraphics[width=0.49\textwidth]{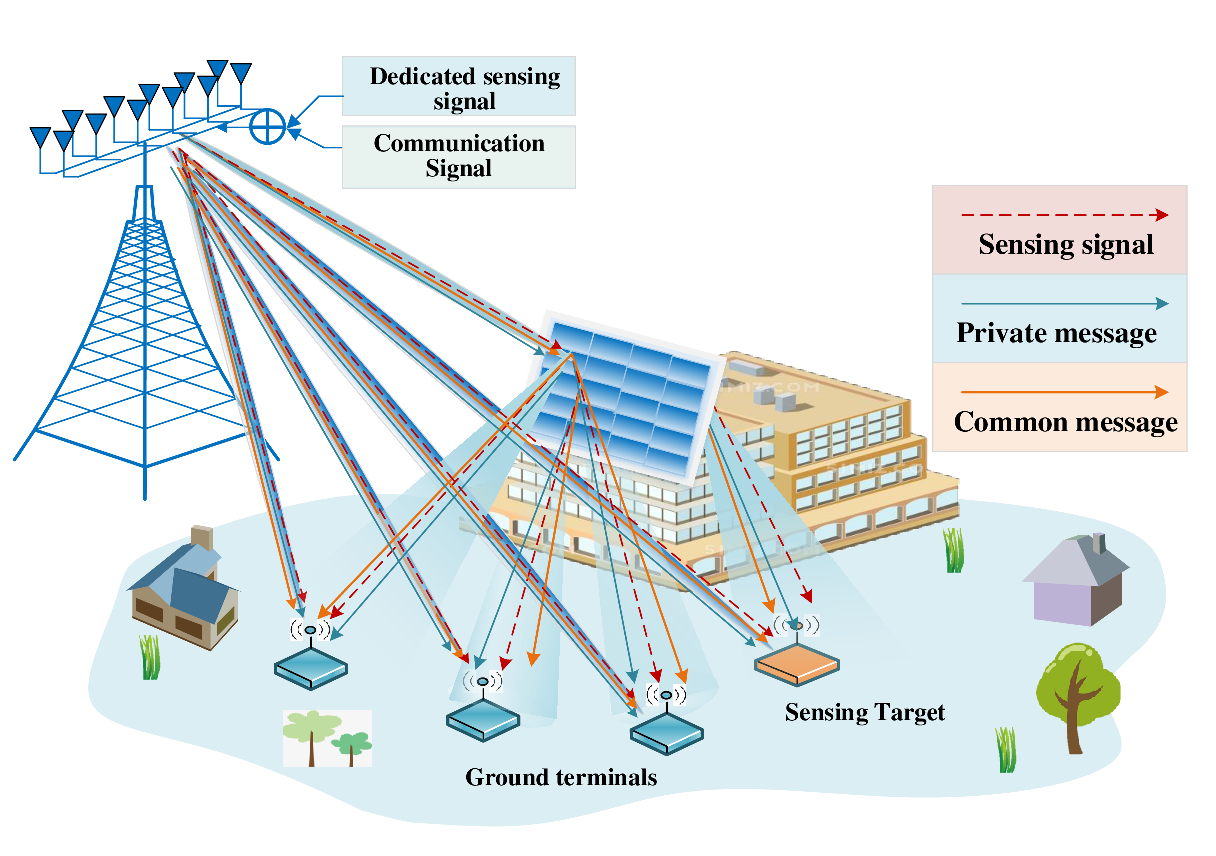}
\caption{STAR-RIS empowered downlink ISAC system with RSMA scheme.}
\label{system_model_ISAC}
\end{figure}
As shown in Fig.~\ref{system_model_ISAC},
a STAR-RIS-enabled downlink ISAC system with
RSMA transmission is examined.
Specifically, the ISAC BS is configured with $N$ transmit antennas and $N$ receive antennas with
half-wavelength spacing,
which are arranged in the form of uniform linear arrays (ULAs).
This BS not only transmits information to multiple GTs
but also is able to detect one target
\footnote{
In this paper,
we focus on a single target.
For scenarios involving multiple targets,
a time division (TD) sensing scheme can be utilized.
By sequentially sensing each target in distinct, orthogonal time slots,
it becomes feasible to utilize all the elements of STAR-RIS to concentrate the beam pattern exclusively towards on a target at any particular moment.
The interference among different targets can be prevented.
The scenario of single-target sensing can be regarded as a special case of multi-target sensing with the TD sensing approach.
}
 through multi-antenna beamforming.
Furthermore,
to overcome the limitations imposed by traditional RISs in terms of half-space coverage,
the STAR-RIS with full-space coverage is employed to coordinate with the BS through a controller
for signal enhancement.
On the other hand,
during downlink transmission,
the RSMA scheme is employed to serve $K$ single-antenna GTs
with $k\in\{1,2,...,K\}$.
The RSMA scheme involves splitting the message transmitted to the $k$-th GT, denoted as $s_k$, into two parts: the common (or public) part, denoted as $s_{c,k}$, and the private part, denoted as $s_{{\rm p},k}$.
Subsequently,
by using a commonly shared codebook,
the common components of all GTs $\{s_{c,1},s_{c,2},...,s_{c,K}\}$ are
combined into a common stream $s_c$,
while the beamforming vector for the common stream $s_c$ is
denoted as $\bm \varpi_c \in \mathbb{C}^{N \times 1}$.
The private part of $k$-th GT is encoded using a private codebook known exclusively by the BS and itself,
and the beamforming vector associated with the private message of $k$-th GT is
denoted as $\bm \varpi_{{\rm p},k} \in \mathbb{C}^{N \times 1}$.
Moreover,
the dedicated radar signal transmitted for sensing the target
is $s_0$ with beamforming vector
denoted as $ \bm \varpi_0 \in \mathbb{C}^{N \times 1}$.

\subsection{Transmission Model}\label{T model}
%Specifically,
%the locations of the ISAC BS is denoted as $q_{\rm B} = [0, 0, 0]$,
%and the locations of $K$ GTs are denoted as $q_{k} = [x_k, y_k, 0]$,
%with $k = \{1,...,K\}$.
%Moreover,
%the location of the sensing target is denoted as $s_t = [t_x, t_y, 0]$.
The STAR-RIS encompasses ${\bm M}$ elements capable of both transmission and reflection.
Each element engages the ES mode to split incoming signals into two segments,
subsequently redirecting them
towards the transmission and reflection spaces to perform their respective communication and sensing tasks.
The transmission (t) and reflection (r) coefficient matrices of STAR-RIS are represented as
\begin{equation}
\begin{aligned}
&\boldsymbol {\Phi }_{p}= \mathrm {diag}\left ({\sqrt {\beta _{1}^{p}} e^{\,j \theta _{1}^{p}}, \sqrt {\beta _{2}^{p}} e^{\,j \theta _{2}^{p}}, \ldots, \sqrt {\beta _{M}^{p}} e^{\,j \theta _{M}^{p}}}\right), \\
&~\forall ~p \in \{\rm t, r\}
\end{aligned}
\end{equation}
where
$\sqrt{\beta_{m}^p} \in [0,1]$
signifies the amplitude and
${\theta_{m}^p} \in [0, 2\pi)$
signifies the phase-shift value
of the $m$-th STAR-RIS element with $ m \in \{1,2,...,M\}$.
To uphold the principle of energy conservation \cite{intro_star_RIS_ISAC_ref1_sdma},
the amplitude adjustments of all STAR-RIS elements should meet specific criteria,
i.e.,
${\beta_{m}^{\rm t}} + {\beta_{m}^{\rm r}}$ = 1.
We consider that the GTs
are stationed on the front face of STAR-RIS,
receiving the reflected signals,
whereas the target is positioned
on the rear side of STAR-RIS,
capturing the transmitted signals.
With both the direct link and STAR-RIS link considered,
the received signal at the $k$-th
GT can be represented as
\begin{equation}
\begin{aligned}
{y_{k}}&=
\left({\boldsymbol  h}_{{\rm b}k}^{H} +{\boldsymbol h}_{{\rm r}k}^{ H}  \bm \Phi_{\rm r}  {\bm H}_{{\rm br}}^{H}  \right) { \bm \varpi_c s_{c}}\\
& + \left({\boldsymbol h}_{{\rm b}k}^{H} +{\boldsymbol h}_{{\rm r}k}^{ H}  \bm \Phi_{\rm r}  \bm{H}_{{\rm br}}^{H} \right) \sum \nolimits_{k=1}^K  \bm \varpi_{{\rm p},k} s_{{\rm p},k}\\
& + \left({\boldsymbol h}_{{\rm b}k}^{H} +{\boldsymbol h}_{{\rm r}k}^{ H}  \bm \Phi_{\rm r}  \bm{H}_{{\rm br}}^{H}  \right) \bm \varpi_0s_0+  n_{k},
% \bf{h_{k}} = \left(\bm{h}_{{\rm B}k} +  {\bm h}_{{\rm R}k} \bm \Phi \bm{h}_{{\rm BR}}  \right)
\end{aligned}
\end{equation}
where ${\boldsymbol h}_{{\rm b}k} \in \mathbb{C}^{N \times 1}$ represents the communication channel from BS to $k$-th GT,
${\boldsymbol h}_{{\rm r}k} \in \mathbb{C}^{M \times 1}$ signifies the communication channel linking STAR-RIS to $k$-th GT,
and $\bm{H}_{{\rm br}} \in \mathbb{C}^{N \times M}$ represents the communication link from BS to STAR-RIS.
Besides,
$\bm \Phi_{\rm r}$ indicates the reflecting coefficient matrix associated with the STAR-RIS,
$n_{k}$ stands for the additive white Gaussian
noise (AWGN) with the power of $\sigma_k^2$ at $k$-th GT.
%The ISAC signals transmitted by the BS
%may experience the distance-dependent
%path loss and small-scale Rician fading,
%and then
%the communication channel from the BS to the $k$-th user can be modeled by
%\begin{equation}
%\begin{aligned}
%\bf{h_{k}} = \left(\bm{h}_{{\rm B}k} +  {\bm h}_{{\rm R}k} \bm \Phi \bm{h}_{{\rm BR}}  \right)
%\end{aligned}
%\end{equation}
For GTs,
they should firstly decode common message $s_c$ by
sharing the commonly code-book among GTs,
where the private message and sensing signal are treated as interference.
Thus, by defining
${\boldsymbol h}_{k} = {\boldsymbol h}_{{\rm b}k} + \bm{H}_{{\rm br}} {\bm \Phi_{\rm r}} {\boldsymbol h}_{{\rm r}k}$,
the achievable rate for GT $k$ to decode common message $s_c$ is expressed as
\begin{equation}
\begin{aligned}
&R_{c,k} = \log_2\left(1+\frac{  \left|  {\boldsymbol h}_{k}^H \bm \varpi_c\right|^2}{\sum \limits_{i=1}^K \left|{\boldsymbol h}_{k}^H  {\bm \varpi_{{\rm p},i}} \right|^2+
\left|  {\boldsymbol h}_{k}^H  \bm \varpi_0\right|^2 +\sigma_k^2}\right).
\label{rc_rate_1}
\end{aligned}
\end{equation}
Subsequently,
the common message is subtracted from the received signal $y_k$ through SIC,
ensuring that private messages are able to be decoded independently, avoiding by interference from the common message.
Therefore,
the attainable rate for GT $k$ to decode its designated private message $s_{{\rm p},k}$ is determined as follows:
\begin{equation}
\begin{aligned}
&R_{{\rm p},k} = \log_2 \left(1+\frac{ \left|{\boldsymbol h}_{k}^H{\bm \varpi_{{\rm p},k}} \right|^2}{
\sum \nolimits_{j\neq k}^K \left|{\boldsymbol h}_{k}^H {\bm \varpi_{{\rm p},j}} \right|^2+
\left|  {\boldsymbol h}_{k}^H  \bm \varpi_0\right|^2 +\sigma_k^2}\right).
\label{rp_rate_1}
\end{aligned}
\end{equation}
Furthermore, to ensure successfully decoding of common messages by all GTs, the achievable rates for each GT should exceed the allocated data rates \cite{intro_RSMA_equation}.
This implies that, for every GT to accurately detect its designated information from the common message, the following constraint should be fulfilled,
i.e.,
\begin{equation}
\begin{aligned}
\sum \nolimits_{k=1}^{K} c_k \le R_c,~ R_c = \min\{R_{c,1},R_{c,2},...,R_{c,K}\},
\label{com_pri_rate_equation}
\end{aligned}
\end{equation}
where $c_k$ represents the actual data rate assigned to $k$-th GT.

\subsection{Sensing Model}
%It is assumed that the number of receiving antennas of ISAC BS is $N$.
The ISAC signals emitted by the BS
initially arrive at the detection target through both the direct link and the refractive path facilitated by STAR-RIS.
Subsequently,
the echo signal,
upon reflection from the target,
traverse similar paths back to the BS.
By defining ${\boldsymbol h}_{\rm bt} \in \mathbb{C}^{N \times 1}$ and ${\boldsymbol h}_{\rm rt} \in \mathbb{C}^{M \times 1}$
as the channel links connecting the BS to target,
and the STAR-RIS to target, respectively,
and $ \bm \Phi_{\rm t}$ as the transmission coefficient matrix associated with STAR-RIS,
the echo signal reflected by the target and $I$ scatters is represented as
\begin{equation}
\begin{aligned}
&{\boldsymbol y}_{\rm bs} =  ({\boldsymbol h}_{\rm bt} + {\bm H}_{\rm br}\bm \Phi_{\rm t}{\boldsymbol h}_{\rm rt})({\boldsymbol h}_{\rm bt} + {\bm H}_{\rm br}\bm \Phi_{\rm t}{\boldsymbol h}_{\rm rt})^H {\bm A}(\theta_0) \times  \\
 & (\sum \nolimits_{k=1}^K  \bm \varpi_{{\rm p},k} s_{{\rm p},k} +{ \bm \varpi_c s_{c}}+\bm \varpi_0s_0 ) + {\sum \nolimits_{i=1}^I {\boldsymbol z}_i}
 + {\boldsymbol n}_s,
\label{eq_1}
\end{aligned}
\end{equation}
% & \sum \nolimits_{i=1}^I \beta_i a_{\rm r}(\theta_i)a_{\rm t}(\theta_i)^H {\bf h}_i {\bf h}_i^H W
where the intended target for detection is located at an angle of $\theta_0$,
and ${\bf A}(\theta_0) = \beta_0 {\boldsymbol a}(\theta_0) {\boldsymbol a}(\theta_0)^H$
with
${\boldsymbol a}(\theta_0) \triangleq [1,e^{j\pi\sin(\theta_0)}, ... , e^{j\pi (N-1)\sin(\theta_0)}]^T$ denoting as the steering vector of the antenna array at the BS
\cite{intro_transmit_array}.
Besides,
$\beta_0 \in \mathbb C$ denotes the complex reflection factor,
${\boldsymbol z}_i \in  \mathbb C^{N\times1}$ represents the undesired
single-independent interference
from $I$ uncorrelated scatters
positioned at angles
$\{\theta_i\}\sum \nolimits_{i=1}^I$
with $\theta_i \neq \theta_0 $ and $ i \in \{1,..., I\}$,
${\boldsymbol n}_s \in \mathbb C^{N\times1}$ denotes the
AWGN at the BS with ${\boldsymbol n}_s \sim \mathcal{CN}(0, \sigma_s^2\bm I_N)$.
By defining
${\boldsymbol h}_{\rm t} = {\boldsymbol h}_{\rm bt} + {\bm H}_{\rm br}\bm \Phi_{\rm t}{\boldsymbol h}_{\rm rt} $,
the output sensing SINR is given by
\begin{equation}
\begin{aligned}
\gamma_{\rm bs} =
\frac{\Vert {\boldsymbol h}_{\rm t} \Vert^2
{\rm Tr} \left(  {\boldsymbol h}_{\rm t} {\boldsymbol h}_{\rm t}^H  {\bf A}(\theta_0) {\bf Q}{\bf A}(\theta_0)^H \right)}
{
{\rm Tr}\left(\sum\nolimits_{i=1}^I  \Vert {\boldsymbol h}_{ i} \Vert^2 {\boldsymbol h}_i {\boldsymbol h}_i^H {\bf A}(\theta_i) {\bf Q}{\bf A}(\theta_i)^H  + \sigma_s^2 {\bf I}_N \right)},
\label{gamma_sense_initial}
\end{aligned}
\end{equation}
where ${\bf Q}$ =  ${ \boldsymbol \varpi}_{c} { \bm\varpi}_{c}^H$ +
$\sum \nolimits_{k=1}^{K}{\bm \varpi_{{\rm p},k}}{\bm \varpi_{{\rm p},k}}^H$ +
${ \boldsymbol \varpi}_{0} { \bm\varpi}_{0}^H $.
Moreover,
${\boldsymbol h}_i$ = ${\boldsymbol h}_{\rm bi} + {\bm H}_{\rm br}\bm \Phi_{\rm t}{\boldsymbol h}_{\rm ri} $,
where ${\boldsymbol h}_{\rm bi}$ and ${\boldsymbol h}_{\rm ri}$
are the communication channel connecting the BS to $i$-th scatter and
the communication link connecting STAR-RIS to $i$-th scatter.
%In addtion,
%$\beta_0$ and $\beta_i$ are the complex amplitude of the target and $i$-th scatter determined by Radar Cross Section (RCS).

\section{Problem formulation and proposed solution}\label{problem_formulation}
\subsection{Problem Formulation}
To maximize the sensing SINR while ensuring the communication rate requirements of all GTs can be guaranteed,
an optimization problem is formulated by simultaneously optimizing the transmit beamforming vectors of the BS $\{\bm \varpi_{c}, \bm \varpi_{{\rm p},k},  \bm \varpi_{0}\}$,
alongside the transmission and reflection beamforming matrices
$\{{\bm \Phi}_{\rm t}, {\bm \Phi}_{\rm r}\}$,
and the rate allocation vector $c_k$ for the common stream.
Specifically, the sensing SINR maximization problem
is mathematically modeled by
\begin{subequations}
\begin{align}
&\textbf{P}_\textbf{A}:
\mathop{\max}
\limits_{\bm \varpi_c, \bm \varpi_{{\rm p},k},\bm \varpi_0,c_k, {\bm \Phi}_{\rm t}, {\bm \Phi}_{\rm r}}
{\kern 1pt} \gamma_{\rm bs} \nonumber \\
{\rm{s.}} {\rm{t.}} &~
{\rm Tr}(\bm \varpi_c \bm \varpi_c^H) + {\rm Tr}(\sum \nolimits_{k=1}^{K} {\bm \varpi}_{{\rm p},k} {\bm  \varpi}_{{\rm p},k}^H) \nonumber \\
& + {\rm Tr}({\bm \varpi}_0 {\bm \varpi}_0^H) \le P_{\max}, \label{cs_1}\\
&~\sum \nolimits_{k=1}^{K} c_k \le R_c, R_c = \min\{R_{c,1},R_{c,2},...,R_{c,K}\},\label{cs_2}\\
&~ c_k + R_{p,k} \ge R_{k}^{\rm th}, \label{cs_3} \\
&~c_k \ge 0, ~\forall ~k \in \{1,...,K\}, \label{cs_33}\\
&~ {\beta_{m}^{\rm t}} + {\beta_{m}^{\rm r}} = 1, ~\forall~ m \in \{1,...,M\}, \label{cs_4}\\
&~ \sqrt{\beta_{m}^p} \in [0,1] , ~ \theta_{m}^p \in [0, 2\pi), ~~p\in \{\rm t,r\},  \label{cs_5}
\end{align}
\end{subequations}
where (\ref{cs_1}) is the transmit power constraint
with $P_{\max}$ being the maximum available power at the BS.
Constraint (\ref{cs_2}) signifies that the total actual rate assigned to the GTs must not exceed the attainable common-stream rate among all GTs.
(\ref{cs_3}) is the minimum rate requirement for each GT with
$R_{k}^{\rm th}$ being the rate threshold of $k$-th GT.
(\ref{cs_4}) signifies that,
owing to passive characteristics of STAR-RIS,
the amplitude responses of all elements are confined by the principle of energy conservation.
(\ref{cs_5}) represents the permissible range of the
transmission coefficients (TCs) and reflection coefficients (RCs)
of STAR-RIS elements, respectively.
It should be noted that the beamforming vectors $\{\bm \varpi_{c}, \bm \varpi_{{\rm p},k}, \bm \varpi_0\}$ and
the TCs and RCs $\{{\bm \Phi}_{\rm t}, {\bm \Phi}_{\rm r}\}$
are multiply coupled together in the optimization objective and in constraints (\ref{cs_2}) and (\ref{cs_3}).
%Additionally,
%the optimization variables pertaining to amplitude adjustment in constraint (\ref{cs_4}) possess a module value of one.
Consequently,
problem $\textbf{P}_\textbf{A}$ is non-convex.
% posing highly challenges in obtaining solutions.
Hence,
we intend to utilize the SDR, MM, and SROCR methods to
devise the successive convex approximation (SCA)-based iterative
algorithm,
which is capable of finding the solution of $\textbf{P}_\textbf{A}$.

\subsection{Proposed Algorithm}
In this section,
we present the SCA-based iterative algorithm aimed at solving the formulated problem.
Initially,
as the matrices of TCs and RCs are coupled with the
variables associated with transmit beamforming and
the actual data rate assigned to each GT,
the initial problem $\textbf{P}_\textbf{A}$ is split into two
sub-problems.
Firstly,
with given initial value of TCs and RCs $\{{\bm \Phi}_{\rm t}^{(l)}, {\bm \Phi}_{\rm r}^{(l)}\}$,
the first sub-problem $\textbf{P}_\textbf{B1}$ w.r.t.
the transmit beamforming vectors and the real data rate allocation
is successively tackled by implementing
Dinkelbach's transform, SDR,
first-order Taylor expansion and variable substitution.
Secondly,
building upon the solution acquired through solving $\textbf{P}_\textbf{B1}$,
the second sub-problem $\textbf{P}_\textbf{B2}$ w.r.t.
the TCs and RCs of STAR-RIS is tackled by
utilizing the SDR,
MM and the SROCR methods.
Finally,
the overall solution for the original problem is
derived by iteratively solving two sub-problems.

To obtain a tractable solution method,
a classic SDR-based approach is applied.
Specifically,
by applying the equivalent transformation
of $ { \bf  W}_c$ = ${  \bm \varpi}_{c} { \bm \varpi}_{c}^H$,
${\bf W}_{{\rm p},k} = {\bm \varpi}_{{\rm p},k} { \bm \varpi}_{{\rm p},k}^H$, $ { \bf  W}_0 = { \bm \varpi}_{0} { \bm \varpi}_{0}^H$,
with given ${\bf H}_k = {\boldsymbol h}_{k}{\boldsymbol h}_{k}^H$
and
$\{{\bm \Phi}_{\rm t}^{(l)}$, ${\bm \Phi}_{\rm r}^{(l)}\}$,
the subproblem $\textbf{P}_\textbf{B1}$ w.r.t.
$\{{ \bf  W}_c,{\bf W}_{{\rm p},k}$,${ \bf  W}_0,c_k\}$
is given by
%\subsubsection{Optimizing $\{{ \bf  W}_c,{\bf W}_{{\rm p},k}$,${ \bf  W}_0,c_k\}$ with given ${\bm \Phi}_{\rm t}^{(l)}$, ${\bm \Phi}_{\rm r}^{(l)}$}
%By defining
%and
\begin{subequations}
\begin{align}
&\textbf{P}_\textbf{B1}:
\mathop{\max}
\limits_{{ \bf  W}_c, {\bf W}_{{\rm p},k},{ \bf  W}_0, c_k }
{\kern 1pt} \gamma_{\rm bs} \nonumber \\
{\rm{s.}} {\rm{t.}} &~
{\rm Tr}({ \bf  W}_c + {\bf W}_{{\rm p},k}+{ \bf  W}_0)\le P_{\max}, \label{ccs_1}\\
&{\rm rank}({ \bf  W}_c) = 1,{\rm rank}({\bf W}_{{\rm p},k}) = 1,{\rm rank}({\bf W}_0) = 1, \\
&{\bf W}_{{\rm p},k} \succeq 0,~
{\bf W}_{c} \succeq 0,~{\bf W}_{0} \succeq 0,\\
&\sum \limits_{k=1}^{K} c_k \le \mathop{\min}
\limits_{\forall k }
\log_2(1+\tfrac{{\rm Tr}( {\bf H}_k {\bf W}_c)}{
\sum \limits_{i=1}^K {\rm Tr}({\bf H}_k ({\bf W}_{{\rm p},i} +
{\bf W}_0))  + \sigma_k^2}), \label{cs_22}\\
& c_k + \log_2(1+\tfrac{{\rm Tr}( {\bf H}_k {\bf W}_{{\rm p},k})}{
\sum \limits_{j\neq k}^K {\rm Tr}({\bf H}_k {\bf W}_{{\rm p},j}  + {\bf H}_k {\bf W}_0) + \sigma_k^2})\ge R_{k}^{\rm th}\label{cs_33}.
\end{align}
\end{subequations}

\newtheorem{theorem}{Theorem}
\begin{theorem}\label{thm:my1}
If the optimized beamforming vectors for
sub-problem ${\mathbf {P}}_{\textbf{B1}}$ is denoted
as $\{\widehat{{\bf W}}_c,
\widehat{{\bf W}}_{p,k}, \widehat{{\bf W}}_0\}$,
there always exists another set of solutions
$\{\overline{{\bf W}}_c,\overline{{\bf W}}_{p,k}, \overline{{\bf W}}_{0}\}$.
It can achieve same system performance not inferior to that of
$\{\widehat{{\bf W}}_c, \widehat{{\bf W}}_{p,k},
\widehat{{\bf W}}_0\}$,
where
\begin{equation}
\begin{aligned}
&\overline{{\bf W}}_c = \widehat{{\bf W}}_c +
\zeta_c \widehat{{\bf W}}_0,~
\overline{{\bf W}}_{p,k} = \widehat{{\bf W}}_{p,k} + \zeta_k \widehat{{\bf W}}_0,\\
&\overline{{\bf W}}_{0} = 0,~~\zeta_c + \sum \nolimits_{k=1}^K \zeta_k = 1.
\label{eq_yrpsnc}
\end{aligned}
\end{equation}
\end{theorem}
\begin{IEEEproof}[Proof]
See Appendix A.
\end{IEEEproof}
From Theorem \ref{thm:my1},
we can observe that
the optimization variable ${\bf W}_{0}$ is unnecessary.
It also reveals that
when the BS only transmits
communication waveforms, it can achieve the same sensing
SINR as when transmitting a combination of communication
and dedicated sensing waveforms,
which is verified by our simulation results in Section \ref{simulation results}.
In other words,
in each iteration,
an optimal solution obtained under the combined waveform assumption can always be equivalently transformed into those corresponding to the communication-only waveform assumption,
and this transformation is independent of the transmission and reflection beamforming design at the STAR-RIS.
Therefore,
the optimization variables of ${\bf W}_{0}$ can be eliminated
to offer a simplified optimization form of problem ${\mathbf {P}}_{\textbf {B1}}$.

However,
the simplified problem is still non-convex.
In the following part,
we apply
Dinkelbach's transform, SDR, first-order Taylor expansion
to deal with non-convex objective and constraints.
Firstly,
the achievable rate for GT $k$ to decode common message
is re-expressed as
\begin{equation}
\begin{aligned}
&R_{c,k} = \log_2\left(1+\frac{{\rm Tr}( {\bf H}_k {\bf W}_c)}{
\sum \nolimits_{i=1}^K {\rm Tr}({\bf H}_k {\bf W}_{{\rm p},i})  + \sigma_k^2}\right).
\label{rate_k_tr}
\end{aligned}
\end{equation}
% {\rm Tr}({\bf H}_k {\bf W}_0)
To address the non-convex constraint (\ref{cs_22}),
we introduce a slack variable ${ a}_k$
and impose additional inequality constraints,
thereby transforming it into the convex one,
i.e.,
\begin{flalign}
\rho_k   {\rm Tr}\left({\bf H}_{k}
({\bf W}_c +
 \sum \nolimits_{i=1}^K {\bf W}_{{\rm p},i} )  \right) +1\ge 2^{a_k},
\label{rate_k_fenzi}
\end{flalign}
and
\begin{equation}
\begin{aligned}
2^{a_k - {(\sum \nolimits_{k=1}^{K} c_k)}} \ge
{ \rho_k  {\rm Tr} \left( {\bf H}_{k} (\sum \nolimits_{i=1}^K{\bf W}_{{\rm p},i}) \right)+1},
\label{rate_k_fenmu}
\end{aligned}
\end{equation}
% + {\bf W}_0
with $\rho_k = \tfrac{1}{\sigma_k^2}$.
Nevertheless,
since constraint (\ref{rate_k_fenmu}) continues to be non-convex,
the first-order Taylor expansion is utilized to tackle it.
Given feasible solution $\{{a_k}^{(l)},{c_k}^{(l)}\}$,
it fulfills the condition that
\begin{equation}
\begin{aligned}
\label{rate_k_fenmu_common}
&2^{{a_k}^{(l)} - \sum \nolimits_{k=1}^{K} {c_k}^{(l)}}
+ \ln2(2^{{a_k}^{(l)} - \sum \nolimits_{k=1}^{K} {c_k}^{(l)}})
(a_k-{a_k}^{(l)}) \\
&- \ln2(2^{{a_k}^{(l)} - \sum \nolimits_{k=1}^{K} { c}_k^{(l)}})
(\sum \nolimits_{k=1}^{K} c_k - \sum \nolimits_{k=1}^{K} {c_k^{(l)}})
 \ge\\
&{ \rho_k  {\rm Tr} \left( {\bf H}_{k} (\sum \nolimits_{i=1}^K{\bf W}_{{\rm p},i}) \right)+1}.
\end{aligned}
\end{equation}
Similarly,
for non-convex constraint (\ref{cs_33}),
by introducing auxiliary variable $b_k$,
we can convert it into the equivalent inequality constraints as follows,
\begin{equation}
\begin{aligned}\label{rate_k_fenz_private}
{ \rho_k  {\rm Tr} \left( {\bf H}_{k} (\sum \nolimits_{i=1}^K{\bf W}_{{\rm p},i}
) \right)+1} \ge
2^{b_k},
\end{aligned}
\end{equation}
and
\begin{equation}
\begin{aligned}
2^{b_k + c_k -R_{\rm th}^k} \ge
{ \rho_k  {\rm Tr} \left( {\bf H}_{k} (\sum \nolimits_{j \neq k}^K{\bf W}_{{\rm p},j}) \right)+1}.
\label{rate_k_private_fenmu}
\end{aligned}
\end{equation}
With the first-order Taylor expansion employed,
the (\ref{rate_k_private_fenmu}) is reexpressed as
\begin{equation}
\begin{aligned}
\label{rate_fenmu_private}
&2^{{b_k}^{(l)}+ {c_k}^{(l)}-R_{\rm th}^k} + \ln2(2^{{b_k^{(l)}}+ {c_k^{(l)}}-R_{\rm th}^k} )(b_k +c_k - {b_k^{(l)}}-{c_k^{(l)}}) \ge \\
& { \rho_k  {\rm Tr} \left( {\bf H}_{k} (\sum \nolimits_{j \neq k}^K{\bf W}_{{\rm p},j}) \right)+1}.
\end{aligned}
\end{equation}
% {\bf W}_0
Moreover,
by defining ${\bf Q} = {\bf W}_c +
 \sum \nolimits_{i=1}^K {\bf W}_{{\rm p},i} $,
${\bf H}_t = {\bf h}_t {\bf h}_t^H$ and
${\bf A}(\theta_i) =
\beta_i {\boldsymbol a}(\theta_i) {\boldsymbol a}(\theta_i)^H $,
${\bf H}_i = {\boldsymbol h}_i {\boldsymbol h}_i^H$,
the numerator and denominator of equation (\ref{gamma_sense_initial})
are rewritten as ${\bf A}_t$ and (${\bf B}_t  +  {\bf I}_N)$, respectively,
where ${\bf A}_t$ and ${\bf B}_t$ are given by
\begin{equation}
\begin{aligned}
&{\bf A}_t = \gamma_0 {\bf H}_t{\bf H}_t^H {\bf A}(\theta_0){\bf Q}{\bf A}(\theta_0)^H,
\label{At_sense_initial}
\end{aligned}
\end{equation}
and
 \begin{equation}
\begin{aligned}
&{\bf B}_t = \sum \nolimits_{i=1}^I \gamma_i {\bf H}_i{\bf H}_i^H {\bf A}(\theta_i){\bf Q}{\bf A}(\theta_i)^H,
\label{Bt_sense_initial}
\end{aligned}
\end{equation}
with $\gamma_i = \frac{1}{\sigma_s^2}$
and $i \in \{0,1,...,I\}$.
To sum up,
the optimization objective $\gamma_{\rm bs}$ can be re-expressed as
\begin{equation}
\begin{aligned}
\gamma_{\rm bs} =  \frac{{\rm Tr}({\bf A}_t)}
{ {\rm Tr}({\bf B}_t   +  {\bf I}_N)}.
\label{frac_objective}
\end{aligned}
\end{equation}
Unfortunately,
the expression in (\ref{frac_objective})
continues to be non-convex
because of the fractional objective and the ambiguity functions w.r.t. ${\bf Q}$.
To resolve these challenges,
the fractional objective is reshaped
through the application of Dinkelbach's transform,
subsequently converting the problem into an explicit form w.r.t. ${\bf Q}$ \cite{dinckbat_method}.
Explicitly,
by incorporating a new auxiliary variable $\omega$,
the objective function is converted into an alternative form,
which is given by
\begin{equation}
\begin{aligned}
\min \limits_{\bf Q} ~ \omega{ {\rm Tr}({\bf B}_t   +  {\bf I}_N)} - {{\rm Tr}({\bf A}_t)},
\end{aligned}
\end{equation}
where $\omega = \frac{ {{\rm Tr}({\bf A}_t^{(l)})}}{{\rm Tr}({\bf B}_t^{(l)}   +  {\bf I}_N)} $.
Furthermore,
for the rank-one constraints of
${\rm rank}({ \bf  W}_c) = 1$ and
${\rm rank}({\bf W}_{{\rm p},k}) = 1$,
the SROCR-based method
is utilized to obtain locally optimal rank-one solutions \cite{SROCR_method}.
The specific step is to restrict the ratio of the largest eigenvalue to the trace of ${{ \bf  W}_c}$ through a flexible parameter $\tau_c$ ranging from 0 to 1,
so that the rank-one constraint can be substituted with
\begin{equation}
\begin{aligned}
{\bf u}_{\rm max}\left( {\bf W}_c\right)^H  {\bf W}_{c}
{\bf u}_{\rm max}\left( {\bf W}_c\right) \ge \tau_c
{\rm Tr}({\bf W}_{c}),
\end{aligned}
\end{equation}
where
\begin{subnumcases}{\label{wt}}
\tau_{c}^{(l)} = \min \left(1,
\tfrac{{\bf e}_{\rm max}\left( {\bf W}_c^{(l)}\right)}
{{\rm Tr}({\bf W}_{c}^{(l)})} + \delta_c^{(l)}\right), \\
\delta_c^{(l)} \in \{0, 1 -\tfrac{{\bf e}_{\rm max}\left({\bf W}_c^{(l)}\right)}
{{\rm Tr}({\bf W}_{c}^{(l)})} \}. \label{dc}
\end{subnumcases}
Besides,
${\bf u}_{\rm max}\left( {\bf W}_c\right)$ is the largest eigenvector of matrix ${\bf W}_c$,
and ${\bf e}_{\rm max}\left({\bf W}_c\right)$ is the largest eigenvalue of matrix ${\bf W}_c$.
With a similar manner,
the other rank-one constraint of
${\rm rank}({\bf W}_{{\rm p},k}) = 1$ can also be converted into processable forms.
After a series of transformations,
the sub-problem ${\mathbf {P}}_{\textbf {B1}}$ can be rewritten as
\begin{subequations}
\begin{align}
& {{\mathbf {P}}_{\textbf{B1-1}}:}
\min \limits_{ \left \{{  {\bf W}_c, {\bf W}_{{\rm p},k}, c_k, a_k, b_k} \right \}}
\omega{ {\rm Tr}({\bf B}_t   +  {\bf I}_N)} - {{\rm Tr}({\bf A}_t)} \nonumber \\
&{\mathrm{ s.t.}}~ (\ref{rate_k_fenzi}), (\ref{rate_k_fenmu_common}), (\ref{rate_k_fenz_private}),(\ref{rate_fenmu_private}), \nonumber\\
&{\rm Tr}( {\bf W}_c + \sum \nolimits_{k=1}^{K} {\bf W}_{{\rm p},k}  ) \le P_{\max},
\label{PA_power}  \\
& {\bf u}_{\rm max}( {\bf W}_c^{(l)})^H  {\bf W}_{c} {\bf u}_{\rm max}( {\bf W}_c^{(l)}) \ge \tau_c^{(l)}
{ \rm Tr}({\bf W}_{c}), \label{rank_c1_beam}\\
& {\bf u}_{\rm max}( {\bf W}_{{\rm p},k}^{(l)})^H
{\bf W}_{{\rm p},k} {\bf u}_{\rm max}( {\bf W}_{{\rm p},k}^{(l)}) \ge \tau_{k}^{(l)}
{\rm Tr}({\bf W}_{{\rm p},k}),  \label{rank_c2_beam} \\
&{\bf W}_{{\rm p},k} \succeq 0,~
{\bf W}_{c} \succeq 0,~
\end{align}
\end{subequations}
where ${\bf W}_c^{(l)}$ and
${\bf W}_{{\rm p},k}^{(l)}$  represent the optimal solutions
of the $l$-th iteration,
(\ref{rank_c1_beam}) and
(\ref{rank_c2_beam}) are
the relaxed convex constraints
for tackling rank-one constraints, respectively.
So far,
${\mathbf {P}}_{\textbf {B1-1}}$ is a convex problem,
which can be solved by using the following Algorithm \ref{alg:pa1}.

\begin{algorithm}
\caption{Iterative algorithm for solving problem ${\mathbf {P}}_{\textbf{B1-1}}$}
\label{alg:pa1}
\begin{algorithmic}[1]
\STATE{Initialize iteration index $l$ = 0, ${a_k}^{(l)}$, ${b_k}^{(l)}$, ${c_k}^{(l)}$, ${\bf W}_c^{(l)}$, ${\bf W}_{{\rm p},k}^{(l)}$, $\tau_c^{(l)}$, $\tau_k^{(l)}$,and calculate $\omega^{(l)}$;}
\STATE{Initialize $\bm \Phi_{\rm t}^{(l)}$ and $\bm \Phi_{\rm r}^{(l)}$};
\STATE{Define initial step sizes:\\~~~~
$ \delta_c^{(l)} \in \{0, 1 -{{\bf e}_{\rm max}({\bf W}_c^{(l)})}/
{{\rm Tr}({\bf W}_{c}^{(l)})} \}$,\\~~~~
$ \delta_k^{(l)} \in \{0, 1 -{{\bf e}_{\rm max}({\bf W}_{{\rm p},k}^{(l)})}/
{{\rm Tr}({\bf W}_{{\rm p},k}^{(l)})} \}$};
\REPEAT
\STATE{Solve problem $\textbf{P}_\textbf{B1-1}$ with $\tau_c^{(l)}$, $\tau_k^{(l)}$, $\omega^{(l)}$, $\bm \Phi_{\rm t}^{(l)}$ and $\bm \Phi_{\rm r}^{(l)}$};
\IF{problem $\textbf{P}_\textbf{B1-1}$ is solvable }
\STATE{Obtain optimal ${\bf W}_c^*$, ${\bf W}_{{\rm p},k}^*$, ${a_k}^*$, ${b_k}^*$, ${c_k}^*$},
\STATE{$\delta_c^{(l+1)}=\delta_c^{(l)}$, $\delta_k^{(l+1)}=\delta_k^{(l)}$},
\ELSE
\STATE{ ${\bf W}_c^{(l+1)}$=${\bf W}_c^{(l)}$, ${\bf W}_{{\rm p},k}^{(l+1)}$=${\bf W}_{{\rm p},k}^{(l)}$,\\
$\delta_c^{(l+1)}=\delta_c^{(l)}/2$, $\delta_k^{(l+1)}=\delta_k^{(l)}/2$,
}
\ENDIF
\STATE{$\tau_{c}^{(l+1)} = \min \left(1,
\tfrac{{\bf e}_{\rm max}\left( {\bf W}_c^{(l+1)}\right)}
{{\rm Tr}({\bf W}_{c}^{(l+1)})} + \delta_c^{(l+1)}\right)$,\\
$\tau_{k}^{(l+1)} = \min \left(1,
\tfrac{{\bf e}_{\rm max}\left( {\bf W}_{{\rm p},k}^{(l+1)}\right)}
{{\rm Tr}({\bf W}_{{\rm p},k}^{(l+1)})} + \delta_k^{(l+1)}\right)$,
}
\UNTIL{ $\lvert 1-\tau_{c}^{(l+1)}\rvert $ $\le$ $\epsilon_{1}$,
$\lvert 1-\tau_{k}^{(l+1)}\rvert $ $\le$ $\epsilon_{1}$
and
the difference between adjacent objective values
for $\textbf{P}_\textbf{B1-1}$ below $\epsilon_{2}$.
}
\end{algorithmic}
\end{algorithm}

Based on the obtained $ \{{\bf W}_{c}^{(l)},{\bf W}_{{\rm p},k}^{(l)},c_k\}$
of $\textbf{P}_\textbf{B1-1}$,
the sub-problem w.r.t.
$\{{\bm \Phi}_{\rm t}, {\bm \Phi}_{\rm r}\}$ will be solved
in the following part.
%\subsubsection{Optimizing $\{{\bm \Phi}_{\rm t}, {\bm \Phi}_{\rm r}\}$ with given $ \{{\bf W}_{c}^{(l)},{\bf W}_{{\rm p},k}^{(l)}$ \} }
Firstly,
as the optimization variables ${\bm \Phi}_{\rm t}$ and $ {\bm \Phi}_{\rm r}$
are comprised in the expression of transmission channel,
we may attempt to extract variable ${\bm \Phi}_{\rm t}$ and ${\bm \Phi}_{\rm r}$ from the channel expression ${\boldsymbol h}_{\rm t}$ and
${\boldsymbol h}_k$, respectively.
By defining ${\boldsymbol \nu}_t = [\sqrt{\beta_{1}^t}e^{j\theta_{1}^t},
\sqrt{\beta_{2}^t}e^{j\theta_{2}^t}, ...,
\sqrt{\beta_{M}^t}e^{j\theta_{M}^t},1]^T$,
the following transformation is obtained, i.e.,
${\boldsymbol h}_{\rm t}= [{\bm H}_{\rm br} {\rm diag}({\boldsymbol h}_{\rm rt}), {\boldsymbol h}_{\rm bt}]{\boldsymbol \nu}_t$.
Therefore,
${\bf A}_t$ shown in (\ref{At_sense_initial}) can be reformulated as
\begin{equation}
\begin{aligned}
&{\bf A}_t = \gamma_t {\boldsymbol \nu}_t^H {\bf A_1} {\boldsymbol \nu}_t {\boldsymbol \nu}_t^H {\bf B_1}{\boldsymbol \nu}_t\\
&=\gamma_t {{\rm Tr} (   {\bf A_1} {\boldsymbol \nu}_t {\boldsymbol \nu}_t^H  {\bf B_1}{\boldsymbol \nu}_t {\boldsymbol \nu}_t^H }) \\
&\overset{(a)}{=} \gamma_t \widehat{{\boldsymbol \nu}}_t^H ({\bf B_1}^T \otimes {\bf A_1} ) \widehat{{\boldsymbol \nu}}_t,
\label{At_sense}
\end{aligned}
\end{equation}
where $\widehat{{\boldsymbol \nu}}_t = {\rm vec}({\boldsymbol \nu}_t {\boldsymbol \nu}_t^H )$,
and step $(a)$ is well-founded due to the application of the identity
${\rm Tr}({\bf EXFX}) = {\rm vec}({\bf X})^H({\bf F}^T\otimes {\bf E}) {\rm vec}(\bf X) $,
${\bf A_1}$ and ${\bf B_1}$ are defined in (\ref{A1B1}).
\begin{figure*}
\begin{equation}
\begin{aligned}
\label{A1B1}
 &{\bf A_1} = \begin{bmatrix}  {\rm diag}^H({\boldsymbol h}_{\rm rt}) {\bm H}_{\rm br}^H
 {\bm H}_{\rm br}  {\rm diag}({\boldsymbol h}_{\rm rt}),& {\rm diag}^H({\boldsymbol h}_{\rm rt}) {\bm H}_{\rm br}^H {\boldsymbol h}_{\rm bt}\\
 {\boldsymbol h}_{\rm bt}^H  {\bm H}_{\rm br}{\rm diag}({\boldsymbol h}_{\rm rt}) ,
 &{\boldsymbol h}_{\rm bt}^H {\boldsymbol h}_{\rm bt} \end{bmatrix} ,\\
& {\bf B_1} = \begin{bmatrix}  {\rm diag}^H({\boldsymbol h}_{\rm rt})  {\bm H}_{\rm br}^H
{\bf Q} {\bm H}_{\rm br}{\rm diag}({\boldsymbol h}_{\rm rt}),&
  {\rm diag}^H({\boldsymbol h}_{\rm rt}) {\bm H}_{\rm br}^H
  {\bf Q} {\boldsymbol h}_{\rm bt}
  \\
 {\boldsymbol h}_{\rm bt}^H {\bf Q}  {\bm H}_{\rm br}{\rm diag}({\boldsymbol h}_{\rm rt})
 ,&{\boldsymbol h}_{\rm bt}^H  {\bf Q}  {\boldsymbol h}_{\rm bt} \end{bmatrix},\\
  &{\bf A_{i1}} = \begin{bmatrix}
 {\rm diag}^H({\boldsymbol h}_{\rm ri}) {\bm H}_{\rm br}^H
 {\bm H}_{\rm br}  {\rm diag}({\boldsymbol h}_{\rm ri}),& {\rm diag}^H({\boldsymbol h}_{\rm ri}) {\bf H}_{\rm br}^H {\boldsymbol h}_{\rm bi}\\
{\boldsymbol h}_{\rm bi}^H  {\bm H}_{\rm br}{\rm diag}({\boldsymbol h}_{\rm ri}) ,
 &{\boldsymbol h}_{\rm bi}^H {\boldsymbol h}_{\rm bi} \end{bmatrix} ,\\
 & {\bf B_{i1}} = \begin{bmatrix}  {\rm diag}^H({\boldsymbol h}_{\rm ri})  {\bm H}_{\rm br}^H
{\bf Q} {\bm H}_{\rm br}{\rm diag}({\boldsymbol h}_{\rm ri}),&
  {\rm diag}^H({\boldsymbol h}_{\rm ri}) {\bm H}_{\rm br}^H
  {\bf Q} {\boldsymbol h}_{\rm bi}
  \\
 {\boldsymbol h}_{\rm bi}^H {\bf Q}  {\bm H}_{\rm br}{\rm diag}({\boldsymbol h}_{\rm ri})
 ,&{\boldsymbol h}_{\rm bi}^H  {\bf Q}  {\boldsymbol h}_{\rm bi} \end{bmatrix}.
% & {\bf Q} =  {\bf W}_c + \sum \nolimits_{k=1}^K  {\bf W}_{p,k}+  {\bf W}_0.
\end{aligned}
\end{equation}
\rule[0.2\baselineskip]{\textwidth}{0.5pt}
\end{figure*}
By using the same derivation method,
${\bf B}_t$ shown in (\ref{Bt_sense_initial})
is given as
\begin{equation}
\begin{aligned}
&{\bf B}_t = \sum \nolimits_{i=1}^I \gamma_i
{\rm Tr} ( {\bf A_{i1}} {\boldsymbol \nu}_t {\boldsymbol \nu}_t^H  {\bf B_{i1}}{\boldsymbol \nu}_t {\boldsymbol \nu}_t^H ) \\
&\overset{(a)}{=} \sum \nolimits_{i=1}^I \gamma_i \widehat{{\boldsymbol \nu}}_t^H ({\bf B_{i1}}^T \otimes {\bf A_{i1}} ) \widehat{{\boldsymbol \nu}}_t,
\label{At_sense}
\end{aligned}
\end{equation}
where ${\bf A_{i1}}$ and ${\bf B_{i1}}$ are also defined in  (\ref{A1B1}).
Therefore, the optimization objective can be re-arranged as
\begin{equation}
\begin{aligned}
&\min  \limits_{{\boldsymbol \nu}_t } ~\widehat{{\boldsymbol \nu}}_t^H \left(   \sum \limits_{i=1}^I \omega\gamma_i  ({\bf B_{i1}}^T \otimes {\bf A_{i1}} ) - \gamma_0({\bf B_1}^T \otimes {\bf A_1} ) \right)  \widehat{{\boldsymbol \nu}}_t.
\label{objectvie_sense_nut}
\end{aligned}
\end{equation}
The existence of the quartic form of optimization variables ${\boldsymbol \nu}_t$  can be noticed in (\ref{objectvie_sense_nut}).
Subsequently,
the MM method is utilized to generate an appropriate surrogate function by leveraging a lower bound defined in (\ref{objectvie_sense_first_order}),
i.e.,
\begin{equation}
\begin{aligned}
\widehat{{\boldsymbol \nu}}_t^H {\bf F}  \widehat{{\boldsymbol \nu}}_t \approx
(\widehat{{\boldsymbol \nu}}_{t}^{(l)})^H ({\bf F} +{\bf F}^H)   \widehat{{\boldsymbol \nu}}_t
-(\widehat{{\boldsymbol \nu}}_{t}^{(l)})^H {\bf F}(\widehat{{\boldsymbol \nu}}_{t}^{(l)}),
\label{objectvie_sense_first_order}
\end{aligned}
\end{equation}
where ${\bf F} = \sum \nolimits_{i=1}^I  \omega\gamma_i  ({\bf B_{i1}}^T \otimes {\bf A_{i1}} ) - \gamma_t({\bf B_1}^T \otimes {\bf A_1} )  $.
However,
(\ref{objectvie_sense_first_order}) is a complex-valued convex function.
Thus,
by defining $\overrightarrow {{{\boldsymbol \nu}}_{t}} = \begin{bmatrix} \Re \left(\widehat{{\boldsymbol \nu}}_t\right) \\ \Im \left(\widehat{{\boldsymbol \nu}}_t\right) \end{bmatrix}$
and $\overrightarrow{\bf F} =
\begin{bmatrix} \Re{(\bf F)},&-\Im {(\bf F)} \\
\Im{(\bf F)},&\Re{(\bf F)}\end{bmatrix}$,
we first convert (\ref{objectvie_sense_first_order}) into a real-valued one,
i.e.,
\begin{equation}
\begin{aligned}
\widehat{{\boldsymbol \nu}}_t^H {\bf F}  \widehat{{\boldsymbol \nu}}_t &=
\overrightarrow{{{\boldsymbol \nu}}_{t}}^T   \overrightarrow{\bf F}  \overrightarrow {{{\boldsymbol \nu}}_{t}} \\
&\approx 2(\overrightarrow{{{\boldsymbol \nu}}}_{t}^{(l)})^T
\overrightarrow{\bf F}  \overrightarrow {{{\boldsymbol \nu}}_{t}} -
(\overrightarrow{{\boldsymbol \nu}}_{t}^{(l)})^T
 \overrightarrow{\bf F}   \overrightarrow{{\boldsymbol \nu}}_{t}^{(l)}.
\label{objectvie_sense_first_order_real_number}
\end{aligned}
\end{equation}
Through the executions of multiple complex mathematical manipulations,
the first term in (\ref{objectvie_sense_first_order_real_number}) can be   expressed as
\begin{equation}
\begin{aligned}
&2(\overrightarrow{{{\boldsymbol \nu}}}_{t}^{(l)})^T   \overrightarrow{\bf F}  \overrightarrow {{{\boldsymbol \nu}}_{t}} = 2\Re({\boldsymbol \psi}^H \widehat{{\boldsymbol \nu}}_t),
\end{aligned}
\end{equation}
where ${\boldsymbol \psi}^H = \begin{bmatrix}
(\overrightarrow{{{\boldsymbol \nu}}}_{t}^{(l)})^T
\overrightarrow{\bf F} \end{bmatrix}_{1:M^2} -
j \begin{bmatrix} (\overrightarrow{{{\boldsymbol \nu}}}_{t}^{(l)})^T   \overrightarrow{\bf F} \end{bmatrix}_{M^2+1:2M^2}$.
Besides,
based on the fact that
$\widehat{{\boldsymbol \nu}}_t = {\rm vec}({\boldsymbol \nu}_t {\boldsymbol \nu}_t^H )$ and ${\rm \bf x^HF^Hx} = {\rm vec}^H(\bf F){\rm vec}(\rm {\bf xx}^H)$,
we have
\begin{equation}
\begin{aligned}
2\Re({\boldsymbol \psi}^H \widehat{{\boldsymbol \nu}}_t)& = 2\Re( {\boldsymbol \nu}_t^H  (\triangle {\boldsymbol \psi} )^H {\boldsymbol \nu}_t)  \\
&={\boldsymbol \nu}_t^H \left( \triangle({\boldsymbol \psi}) + (\triangle({\boldsymbol \psi}))^H  \right){\boldsymbol \nu}_t,
\end{aligned}
\end{equation}
where $\triangle (\cdot) $
denotes the inverse operation of matrix vectorization $\rm vec(\cdot)$.
Actually,
after a series of mathematical transformations,
we have $\triangle(\boldsymbol \psi) = \sum \nolimits_{i=1}^I  \omega\gamma_i
( {\bf B_{i1}} {\boldsymbol \nu}_t^{(l)} ({\boldsymbol \nu}_t^{(l)})^H  {\bf A_{i1}} ) -
\gamma_0 (   {\bf B_1} {\boldsymbol \nu}_t^{(l)} ({\boldsymbol \nu}_t^{(l)})^H {\bf A_1} ) $.
Up to this point,
the objective function in (\ref{objectvie_sense_nut}) can be equivalently rewritten as
\begin{equation}
\begin{aligned}
&\min  \limits_{\boldsymbol \nu_t } ~{\boldsymbol \nu}_t^H \left( \triangle({\boldsymbol \psi}) + (\triangle({\boldsymbol \psi}))^H  \right){\boldsymbol \nu}_t -(\overrightarrow{\boldsymbol \nu}_{t}^{(l)})^T
 \overrightarrow{\bf F}   \overrightarrow{\boldsymbol \nu}_{t}^{(l)}.
\label{objectvie_sense_final}
\end{aligned}
\end{equation}
Furthermore,
by defining ${\bf V}_{\rm t} = { \boldsymbol \nu}_t {\boldsymbol \nu}_t^H $ and
${\bf V}_{\rm r} = {\boldsymbol \nu}_r {\boldsymbol \nu}_r^H $ with ${\rm rank}({\bf V}_{\rm t}) = 1$
and ${\rm rank}({\bf V}_{\rm r}) = 1$,
constraint (\ref{cs_22})
can be re-written as (\ref{rate_com_fenzi_vr}),
and constraint (\ref{cs_33}) can be rewritten as (\ref{rate_private_fenzi_vr}),
which are given by
\begin{flalign}
&\rho_k   {\rm Tr} ({\bf M}_{Q1}{\bf V}_{\rm r}) +1\ge
2^{ {(\sum \nolimits_{k=1}^{K} c_k)}} \left(\rho_k\rm  Tr({\bf M}_{Q2}{\bf V}_r) + 1\right),
\label{rate_com_fenzi_vr}
\end{flalign}
and
\begin{flalign}
&\rho_k   {\rm Tr}({\bf M}_{Q2}{\bf V}_{\rm r}) +1\ge
2^{ R_{\rm th}^k- c_k} \left(\rho_k  {\rm Tr} ({\bf M}_{Q3}{\bf V}_{\rm r}) + 1\right),
\label{rate_private_fenzi_vr}
\end{flalign}
where ${\bf M}_{Q1}$, ${\bf M}_{Q2}$ and ${\bf M}_{Q3}$ are shown in (\ref{MQ_123}).
\begin{figure*}
\begin{equation}
\begin{aligned}
\label{MQ_123}
 &{\bf M}_{Q1} = \begin{bmatrix}  {\rm diag}^H({\boldsymbol h}_{{\rm r}k}) {\bm H}_{\rm br}^H{\bf Q_1} {\bm H}_{\rm br}{\rm diag}({\boldsymbol h}_{{\rm r}k}),&
 {\rm diag}^H({\boldsymbol h}_{{\rm r}k}) {\bm H}_{\rm br}^H{\bf Q_1} \bm{h}_{{\rm b}k}\\
 {\boldsymbol h}_{{\rm b}k}^H{\bf Q_1}{\bm H}_{\rm br}{\rm diag}({\boldsymbol h}_{{\rm r}k}),&
  {\boldsymbol h}_{{\rm b}k}^H{\bf Q_1} {\boldsymbol h}_{{\rm b}k}
 \end{bmatrix},
 {\bf Q_1}= {\bf W}_c +
 \sum \nolimits_{i=1}^K {\bf W}_{{\rm p},i} \\
  &{\bf M}_{Q2} = \begin{bmatrix}  {\rm diag}^H({\boldsymbol h}_{{\rm r}k}) {\bm H}_{\rm br}^H{\bf Q_2} {\bm H}_{\rm br}{\rm diag}({\boldsymbol h}_{{\rm r}k}),&
 {\rm diag}^H({\bm h}_{{\rm r}k}) {\bm H}_{\rm br}^H{\bf Q_2} {\boldsymbol h}_{{\rm b}k}\\
 {\boldsymbol h}_{{\rm b}k}^H{\bf Q_2} {\bm H}_{\rm br}{\rm diag}({\bm h}_{{\rm r}k}),&
  {\boldsymbol h}_{{\rm b}k}^H{\bf Q_2} {\boldsymbol h}_{{\rm b}k}
 \end{bmatrix},
 {\bf Q_2}=  \sum \nolimits_{i=1}^K {\bf W}_{{\rm p},i} \\
 & {\bf M}_{Q3} = \begin{bmatrix}  {\rm diag}^H({\boldsymbol h}_{{\rm r}k}) {\bm H}_{\rm br}^H{\bf Q_3} {\bm H}_{\rm br}{\rm diag}({\boldsymbol h}_{{\rm r}k}),&
 {\rm diag}^H({\bm h}_{{\rm r}k}) {\bm H}_{\rm br}^H{\bf Q_3} \bm{h}_{{\rm b}k}\\
 {\boldsymbol h}_{{\rm b}k}^H{\bf Q_3} {\bm H}_{\rm br}{\rm diag}({\bm h}_{{\rm r}k}),&
  {\boldsymbol h}_{{\rm b}k}^H{\bf Q_3} {\boldsymbol h}_{{\rm b}k}
 \end{bmatrix},
 {\bf Q_3}=  \sum \nolimits_{j\neq k}^K {\bf W}_{{\rm p},j}
\end{aligned}
\end{equation}
\rule[0.2\baselineskip]{\textwidth}{0.5pt}
\end{figure*}
Therefore,
the optimization problem
related to optimization variables ${\bf V}_{\rm r}$ and ${\bf V}_{\rm t}$ can be
represented as
\begin{subequations}
\begin{align}
& {{\mathbf {P}}_{\textbf {B2}}:}
\min \limits_{ \left \{ {\bf V}_{\rm t},{\bf V}_{\rm r} \right \}}
{\rm Tr} \left({\bf V}_{\rm t} \left( \triangle(\boldsymbol \psi) + (\triangle(\boldsymbol \psi))^H  \right)\right)\nonumber \\
&{\mathrm{ s.t.}}~ (\ref{rate_com_fenzi_vr}),
(\ref{rate_private_fenzi_vr}),  {\bf V}_{\rm t} \succeq 0, {\bf V}_{\rm r} \succeq 0 ,\nonumber\\
& \begin{bmatrix} {\bf V}_{\rm r}\end{bmatrix}_{m,m} +\begin{bmatrix} {\bf V}_{\rm t} \end{bmatrix}_{m,m} == 1,   \\
& 0 \le  \begin{bmatrix} {\bf V}_{\rm r}\end{bmatrix}_{m,m} \le 1,~
\begin{bmatrix} {\bf V}_{\rm r}\end{bmatrix}_{M,M} = 1,\\
& 0 \le  \begin{bmatrix} {\bf V}_{\rm t}\end{bmatrix}_{m,m} \le 1,~
\begin{bmatrix} {\bf V}_{\rm t}\end{bmatrix}_{M,M} = 1, \\
& {\bf u}_{\rm max}( {\bf V}_{\rm t}^{(l)})^H  {\bf V}_{\rm t} {\bf u}_{\rm max}( {\bf V}_{\rm t}^{(l)}) \ge \tau_{t}
{\rm Tr}({\bf V}_{\rm t}),  \label{rank_c2} \\
& {\bf u}_{\rm max}( {\bf V}_{\rm r}^{(l)})^H  {\bf V}_{\rm r} {\bf u}_{\rm max}( {\bf V}_{\rm r}^{(l)}) \ge \tau_{r}
{\rm Tr}({\bf V}_{\rm r}), \label{rank_c3}
\end{align}
\end{subequations}
where ${\bf V}_{\rm t}^{(l)}$ and ${\bf V}_{\rm r}^{(l)}$ represent the optimal solution
obtained from the $l$-th iteration.
To this end,
problem ${\mathbf {P}}_{\mathbf {B2}}$ has been transformed into a convex one,
making it solvable through Algorithm \ref{alg:pa2}.
\begin{algorithm}
\caption{Algorithm for solving ${\mathbf {P}}_{\textbf {B2}}$}
\label{alg:pa2}
\begin{algorithmic}[1]
\STATE{Obtain the optimal solution of ${\bf W}_c$, ${\bf W}_{{\rm p},k}$ and
$c_k$ by solving problem $\textbf{P}_\textbf{B1-1}$}
\STATE{Define initial step sizes:\\~~~~
$ \delta_t^{(l)} \in \{0, 1 -{{\bf e}_{\rm max}({\bf V}_{\rm t}^{(l)})}/
{{\rm Tr}({\bf V}_{\rm t}^{(l)})} \}$,\\~~~~
$ \delta_r^{(l)} \in \{0, 1 -{{\bf e}_{\rm max}({\bf V}_{\rm r}^{(l)})}/
{{\rm Tr}({\bf V}_{\rm r}^{(l)})} \}$.
}
\REPEAT
\STATE{Solve problem ${\mathbf {P}}_{\textbf {B2}}$ with given $\tau_t^{(l)}$, $\tau_r^{(l)}$, $\bm \Phi_{\rm t}^{(l)}$, $\bm \Phi_{\rm r}^{(l)}$};
\IF{problem $\textbf{P}_\textbf{B2}$ is solvable }
\STATE{Obtain optimal ${\bf V}_{\rm t}^*$, ${\bf V}_{\rm r}^*$},
\STATE{$\delta_t^{(l+1)}=\delta_t^{(l)}$, $\delta_r^{(l+1)}=\delta_r^{(l)}$},
\ELSE
\STATE{ ${\bf V}_{\rm t}^{(l+1)}$=${\bf V}_{\rm t}^{(l)}$, ${\bf V}_{\rm r}^{(l+1)}$=${\bf V}_{\rm r}^{(l)}$,\\
$\delta_t^{(l+1)}=\delta_t^{(l)}/2$, $\delta_r^{(l+1)}=\delta_r^{(l)}/2$,
}
\ENDIF
\STATE{$\tau_{t}^{(l+1)} = \min \left(1,
\tfrac{{\bf e}_{\rm max}\left( {\bf V}_{\rm t}^{(l+1)}\right)}
{{\rm Tr}({\bf V}_{\rm t}^{(l+1)})} + \delta_t^{(l+1)}\right)$,\\
$\tau_{r}^{(l+1)} = \min \left(1,
\tfrac{{\bf e}_{\rm max}\left( {\bf V}_{\rm r}^{(l+1)}\right)}
{{\rm Tr}({\bf V}_{\rm r}^{(l+1)})} + \delta_r^{(l+1)}\right)$,\\
}
\UNTIL{ $\lvert 1-\tau_{t}^{(l+1)}\rvert $ $\le$ $\epsilon_{1}$,
$\lvert 1-\tau_{r}^{(l+1)}\rvert $ $\le$ $\epsilon_{1}$,
and the difference between adjacent objective function values is less than threshold $\epsilon_{2}$.
}
\end{algorithmic}
\end{algorithm}

\begin{figure*}[t]
 \centering
\includegraphics[width=0.75\textwidth]{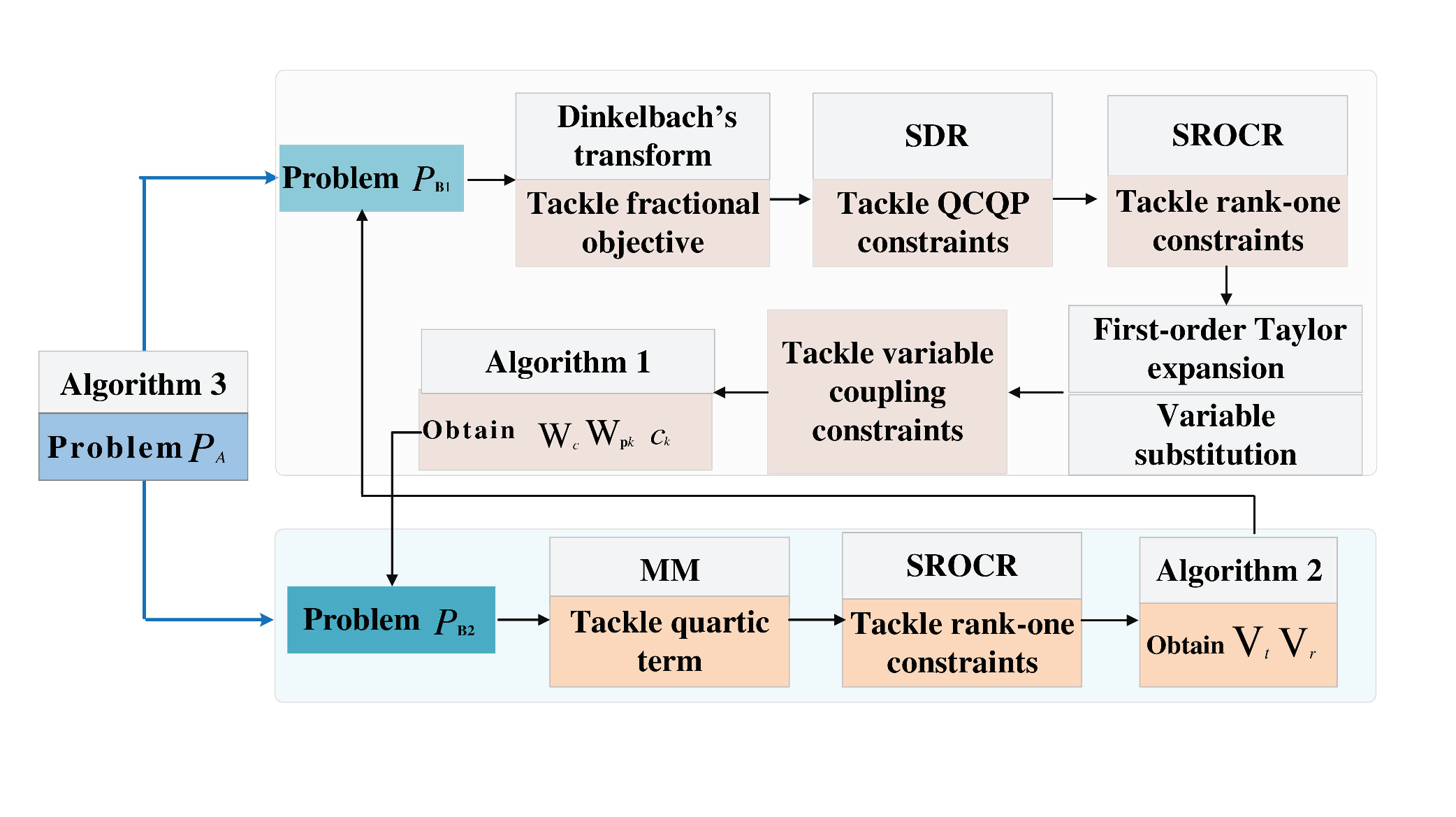}
\caption{ Flowchart of the proposed algorithm.}
\label{flow_chart}
\end{figure*}

By decomposing problem $\textbf{P}_\textbf{A}$ into more tractable sub-problems
$\textbf{P}_\textbf{B1-1}$ and $\textbf{P}_\textbf{B2}$,
we can tackle them independently using Algorithm \ref{alg:pa1} and Algorithm \ref{alg:pa2},
respectively.
Through iterative refinement of both
$\textbf{P}_\textbf{B1-1}$ and $\textbf{P}_\textbf{B2}$,
the optimization variables
$\{{\bf W}_c, {\bf W}_{{\rm p},k}, c_k\}$ and $\{{\bf V}_{\rm t}, {\bf V}_{\rm r}\}$
undergo gradual update with the SCA technique.
The entire iterative algorithm for addressing $\textbf{P}_\textbf{A}$
is elaborated in Algorithm \ref{alg:two-stage},
as well as in Fig.~\ref{flow_chart}.
When the difference between two consecutive solutions obtained while solving $\textbf{P}_\textbf{A}$
achieves the convergence threshold,
the system will output the desired solutions.

\begin{algorithm}
\caption{SCA-based iterative algorithm for solving $\textbf{P}_\textbf{A}$}
\label{alg:two-stage}
\begin{algorithmic}[1]
\STATE{Initialize iteration index $n$ = 1, $\omega^{(n)}$};
\REPEAT
\STATE{
Solve $\textbf{P}_\textbf{B1-1}$ based on Algorithm \ref{alg:pa1},
update ${\bf W}_c$, ${\bf W}_{{\rm p},k}$, $c_k$;
}
\STATE{
Solve $\textbf{P}_\textbf{B2}$ based on Algorithm \ref{alg:pa2},
update ${\bf V}_{\rm t}$, ${\bf V}_{\rm r}$;}
\STATE{Update iteration index $n$ = $n$ + 1, $\omega^{(n+1)}$};
\UNTIL{$\tfrac{\left|\omega^{(n+1)} - \omega^{(n)}\right|}{\omega^{(n)}} \le \varepsilon$ is satisfied}.
\STATE{Obtain the desirable solutions: ${\bf W}_c^*$, ${\bf W}_{{\rm p},k}^*$, $c_k^*$,~$a_k^*$,~ $b_k^*$, ~${\bf V}_{\rm t}^*$, ~${\bf V}_{\rm r}^*$.}
\end{algorithmic}
\end{algorithm}
\subsection {Complexity analysis of proposed algorithm}\label{complexity_analysis_2}
Based on the interior point method \cite{ref_complexity_analysis},
both ${\mathbf{P}}_{\textbf {B1-1}}$ and
${\mathbf {P}}_{\textbf {B2}}$ involve linear matrix inequality (LMI) constraints.
Regarding ${\mathbf{P}}_{\textbf {B1-1}}$,
it comprises $(5K+3)$ one-dimensional LMI constraints
and $(K+2)$ LMI constraints of dimension $N$.
Additionally, it encompasses 4$K$ optimization variables.
Consequently,
the complexity of solving
${\mathbf{P}}_{\textbf {B1-1}}$ can be assessed as
$O( n_1 \cdot \sqrt{6K+5}\cdot ((K+2)(N^3+n_1N^2)+5K+5Kn_1+3n_1+3+ n_1^2))$ with $ n_1 = O(4K) $.
Regarding ${\mathbf{P}}_{\textbf {B2}}$
involving 2 optimization variables,
there are $(2K+3M+2)$ one-dimensional LMI constraints,
and $2$ LMI constraints of dimension $M$.
Consequently,
the complexity of solving
${\mathbf{P}}_{\textbf {B2}}$ is assessed as
$O( n_2 \cdot \sqrt{2K+3M+4)}\cdot (2M^3+2M^2n_2+2Kn_2+2K+3Mn_2+3M+2n_2+ n_2^2))$ with $ n_2 = O(1) $.
To reveal the complexity of proposed algorithm in a clearer way,
by defining $M = \iota_1 K$ and $N = \iota_1 K$ with $\iota_1 = 1$,
the complexities involved in solving the primary problem $\textbf{P}_\textbf{A}$
are outlined in Table \ref{table_complexity}.
%By the same way,
%the complexities of other benchmark schemes are also calculated and listed
%in Table \ref{table_complexity}.

\begin{table*}[!htbp]
\renewcommand{\arraystretch}{1}
	\centering
	\caption{Computational complexity of proposed
 algorithm}
	\begin{tabular}{|c|c|c|}
		\hline
		System design            &   Complexity Order               \\ \hline
STAR-RIS aided ISAC with RSMA   & \makecell[c]{$I_1O\left( n_1 \cdot \sqrt{6K+5}\cdot ((K+2)(N^3+n_1N^2)+5K+5Kn_1+3n_1+3+ n_1^2)\right)$ + \\$I_2O\left( n_2 \cdot \sqrt{2K+3M+4)}\cdot (2M^3+2M^2n_2+2Kn_2+2K+3Mn_2+3M+2n_2+ n_2^2)\right ) $$\approx K^{\frac{11}{2}}$ }\\
  \hline
STAR-RIS aided ISAC with NOMA   & \makecell[c]{
$I_3 O\left( n_3 \cdot \sqrt{k^2+3K+4}\cdot ((K+2)(N^3+n_3N^2)+K^2(n_3+1)+ 3Kn_3+ n_3^2)\right)$ + \\ $I_4 O\left( n_4 \cdot \sqrt{K^2+3M+K+3}\cdot (2M^3+2M^2n_4+  (K^2+3M+K+1)(n_4+1)+ n_4^2)\right) \approx K^6 $ }\\
\hline
STAR-RIS aided ISAC with SDMA   & \makecell[c]{$I_5O\left( n_5 \cdot \sqrt{3K+5}\cdot ((K+2)(N^3+n_5N^2)+2K+2Kn_5+3n_5+3+ n_5^2)\right)$ + \\
$I_6O\left( n_6 \cdot \sqrt{K+3M+4)}\cdot (2M^3+2M^2n_6+Kn_6+K+3Mn_6+3M+2n_6+ n_6^2)\right ) $$\approx K^{\frac{11}{2}}$ }\\
\hline
%		 \multicolumn{2}{|c|}{~~~~~~~~~~~~~~$n_1 = O(4MN +5N+ KMN + KN)$,~~~~~ $n_2 = O(4N)$}\\
% \hline
	\end{tabular}
\label{table_complexity}
\end{table*}

\subsection{Convergence Analysis}
Next,
we evaluate the convergence performance
of the proposed Algorithm \ref{alg:two-stage}.
Let $\gamma_{\rm bs}( \{{ \bf  W}_c^{(n)}, {\bf W}_{{\rm p},k}^{(n)}, c_k^{(n)} \})$
denotes the objective value of problem ${\mathbf {P}}_{\textbf {B1-1}}$ obtained
by Algorithm \ref{alg:pa1}.
Subsequently,
with given $\{{\bm \Phi}_{\rm t}^{(n)}$, ${\bm \Phi}_{\rm r}^{(n)}\}$,
we have
\begin{flalign}
\gamma( \{{ \bf  W}_c^{(n)}, {\bf W}_{{\rm p},k}^{(n)}, c_k^{(n)} \})
\overset{(e)} {\ge} \gamma( \{{ \bf  W}_c^{(n+1)}, {\bf W}_{{\rm p},k}^{(n+1)}, c_k^{(n+1)} \}),
\end{flalign}
where step (e) arises from
the tightness of the first-order Taylor expansion of local points.
As the objective function $\gamma{\rm bs}( \{{ \bf  W}_c, {\bf W}_{{\rm p},k}, c_k \})$
is continuous and the set of feasible solutions is compact,
the objective function $\gamma{\rm bs}( \{{ \bf  W}_c, {\bf W}_{{\rm p},k}, c_k \})$ is
non-increasing,
there exits a lower bound of objective value $\gamma{\rm bs}$
\cite{intro_star_RIS_ISAC_ref2_sdma}\cite{intro_RSMA_ISAC_RIS_ref1}.
Likewise,
with given $\{ { \bf  W}_c^{(n+1)}, {\bf W}_{{\rm p},k}^{(n+1)}, c_k^{(n+1)} \}$
we have
\begin{flalign}
& \gamma( \{{\bm \Phi}_{\rm t}^{(n)}, {\bm \Phi}_{\rm r}^{(n)}\} )
\overset{(f)} {\ge} \gamma(\{{\bm \Phi}_{\rm t}^{(n+1)},{\bm \Phi}_{\rm r}^{(n+1)}\}),
\end{flalign}
where step (f) holds
since the constructed surrogate function in
${\mathbf {P}}_{\textbf {B2}}$ is tight \cite{intro_transmit_array},
and the objective function in ${\mathbf {P}}_{\textbf {B2}}$ is a lower bound of
that in ${\mathbf {P}}_{\mathbf {A}}$.
Therefore,
we have
$\gamma( \{{ \bf  W}_c^{(n)}, {\bf W}_{{\rm p},k}^{(n)}, c_k^{(n)},{\bm \Phi}_{\rm t}^{(n)}, {\bm \Phi}_{\rm r}^{(n)}\} )$
 ${\ge}$ $ \gamma(\{{ \bf  W}_c^{(n+1)}, {\bf W}_{{\rm p},k}^{(n+1)}, c_k^{(n+1)},{\bm \Phi}_{\rm t}^{(n+1)},{\bm \Phi}_{\rm r}^{(n+1)}\})$,
the proposed Algorithm \ref{alg:two-stage} can be guaranteed
to converge over certain iterations.

\section{Numerical results and analysis}\label{simulation results}
This section showcases the simulation results derived from the proposed algorithm. Furthermore,
it offers a comparative analysis in terms of the system performance of our proposed algorithm and other benchmark schemes across diverse system parameters.

\subsection{Simulation Settings}
In the studied system,
%four communication GTs are deployed
%with their respective positions
%designated as follows:
%$q_1$ = [-65,50], $q_2$
%= [-42,40], $q_3$ = [-30,30], $q_4$ = [-20,30].
%Besides,
%the position of sensing target
%is $q_s$ = [60,60],
%and the position of the deployed STAR-RIS is
%$q_r$ = [65,21,10].
%Alternatively,
the communication channels represented by
${\boldsymbol h}_{{\rm b}k}$,
${\boldsymbol h}_{{\rm r}k}$, and $\bm{H}_{{\rm br}}$
are modeled by the combination of distance-dependent path-loss fading and small-scale Rician fading characteristics \cite{intro_path_loss},
i.e.,
\begin{flalign}
& {\boldsymbol h}_{\rm link} = \sqrt{\beta_{\rm link}} \tilde{{\boldsymbol h}}_{\rm link },~~{\rm link }= \{ {{\rm b}k}, {{\rm r}k}, {{\rm br}}\}
\end{flalign}
where $\beta_{\rm link}$ is the path-loss, which is modeled as
$\beta_{\rm link} = \frac{\iota_0}{{\rm d_{link}}^{\alpha_{\rm link}}}$
with $\iota_0$ being the channel power gain of unit reference distance.
Here, ${\rm d_{link}}$ and ${\alpha_{\rm link}}$ correspond to the distance and path loss exponent associated with the transmission link.
% Specifically, the path-loss exponents from the BS to the STAR-RIS, from the BS/STAR-RIS to the GTs, and from the BS/STAR-RIS to the target are set as 2.5, 2.5/2.2, and 3/2.3, respectively.
Furthermore,
$\tilde{\bm h}_{\rm link }$
follows the Rician distribution with Rician factor $K_f$ = 6dB.
%which is generated by
%\begin{flalign}
%& \tilde{{\bf h}}_{\rm link }= \sqrt{\frac{K_f}{1+K_f}}  \tilde{\bf H}_{\rm link}^{\rm LoS} + \sqrt{\frac{1}{1+K_f}}  \tilde{\bf H}_{\rm link}^{\rm NLoS}
%\end{flalign}
%with $K_f$ being the Rician factor.
%In addition,
%$\tilde{\bf H}_{\rm link}^{\rm LoS}$
%denote the deterministic component of the communication link,
%and $\tilde{\bf H}_{\rm link}^{\rm NLoS} \sim \mathcal {CN} (0,1)$ follow the Rayleigh distribution.
The simulation parameter settings
in this paper refer to the works
\cite{intro_star_RIS_ISAC_ref2_sdma},
\cite{intro_RSMA_ISAC_ref1},
\cite{intro_RSMA_ISAC_ref2},
and the specific
parameter configurations are outlined in Table \ref{table_parameters}.

\begin{table}[htb]
\centering
\caption{Parameter Values}
\begin{tabular}{l|c|c}
  \hline
  \hline
  Description & Parameters & Values \\
  \hline
  \hline
	   Number of GTs  &    $K$     &  4  \\
	   Number of STAR-RIS elements   &  $M$     & 64    \\
    Number of antennas of BS  &  $N$     & 6   \\
	   Rate thresholds of GTs   &     $R_{\rm th}$     &\{5, 5, 5, 5\}    \\
    % Position of GTs  &     $q_{k}$     & [10,20;20 30;60 5; 40 4]           \\
       Position of target  &     $s_{t}$     & [89 36 0]           \\
       Position of STAR-RIS  &     $q_{r}$     & [10 90 10]           \\
       The available power of BS   &     $P_{\rm max}$     & 10W                 \\
       Noise power at the GTs  &  $\sigma_{k}^2$   & $10^{-10}$ Watt       \\
       Noise power at the target  &  $\sigma_{s}^2$   & $10^{-12}$ Watt       \\
       Reference channel power gain  &   $\iota_0$     & -15 dB                \\
       Path-loss exponent of BS-GT link  &   $\alpha_{{\rm b}k}$     & 2.7       \\
       Path-loss exponent of BS-target link   &   $\alpha_{{\rm b}s}$     & 2.6       \\
       Path-loss exponent of START-RIS link   &   $\alpha_{r}$     & 2.8       \\
       Rician fading factor  &   $K_f$     & 6dB      \\
      Convergence accuracy  &    $\xi_1,\xi_2,\xi_3$  & $10^{-5}$  \\
       \hline
\end{tabular}
\label{table_parameters}
\end{table}

\subsection{Benchmark Schemes}
To confirm the superiority and  effectiveness of
the considered STAR-RIS-enabled ISAC system
incorporating the RSMA-based approach,
the following
benchmark schemes are introduced for comparison.

\begin{itemize}
\item
\textbf{STAR-RIS: RSMA-N-sensing scheme}:
In this benchmark,
the STAR-RIS aided ISAC system incorporating the RSMA-based approach is investigated,
without taking into account specially designed sensing signals
\cite{intro_star_RIS_ISAC_ref2_sdma}.

\item
\textbf{STAR-RIS: NOMA scheme}:
In this benchmark,
the STAR-RIS-aided ISAC system that employs the NOMA approach is examined,
where the decoding order of SIC is determined by the channel gains of GTs.
Specifically,
the GTs with stronger channel gain should first decode the
GTs' signals with poorer channel conditions,
and then decode their own messages.

\item
\textbf{STAR-RIS: SDMA scheme}:
In this benchmark,
the STAR-RIS aided ISAC system incorporating the SDMA-based scheme is evaluated,
where each GT directly decodes its intended message,
while
disregarding messages designated for other GTs
and perceiving them as interference.

\item
\textbf{Traditional RIS: RSMA scheme}:
In this benchmark,
two RISs, one with transmission capability and the other with reflection capability, are collaboratively deployed to provide full spatial coverage.
It is assumed that their positions coincide with that of the STAR-RIS.
Besides,
to facilitate equitable comparisons,
we assume that every traditional RIS is composed of $M$/2 elements,
given that $M$ is an even number.
It is worth mentioning that the associated optimization problem related to traditional RISs is able to be addressed by setting
$\bm \Phi_{\rm t} =
{\rm diag}\left(e^{j\theta_{1}^t},
e^{j\theta_{2}^t}, ...,
e^{j\theta_{M/2}^t},0,...,0\right)
$
and
$\bm \Phi_{\rm r} =
{\rm diag}\left(0,...,0, e^{j\theta_{M/2+1}^r},...,
e^{j\theta_{M}^r}\right)
$ \cite{intro_NOMA_star_ris}.

\item
\textbf{Random RIS: RSMA scheme}:
In this benchmark,
the TCs and RCs of STAR-RIS elements
are set randomly,
and only the transmit beamforming at the BS is optimized.

\item
\textbf{Non-RIS: RSMA scheme}:
In this benchmark,
the STAR-RIS is absent,
and the base station solely carries out dual functionalities via its direct links to GTs and the intended target.

\end{itemize}

\subsection{Simulation Analysis}
\begin{figure}[t]
 \centering
\includegraphics[width=0.48\textwidth]{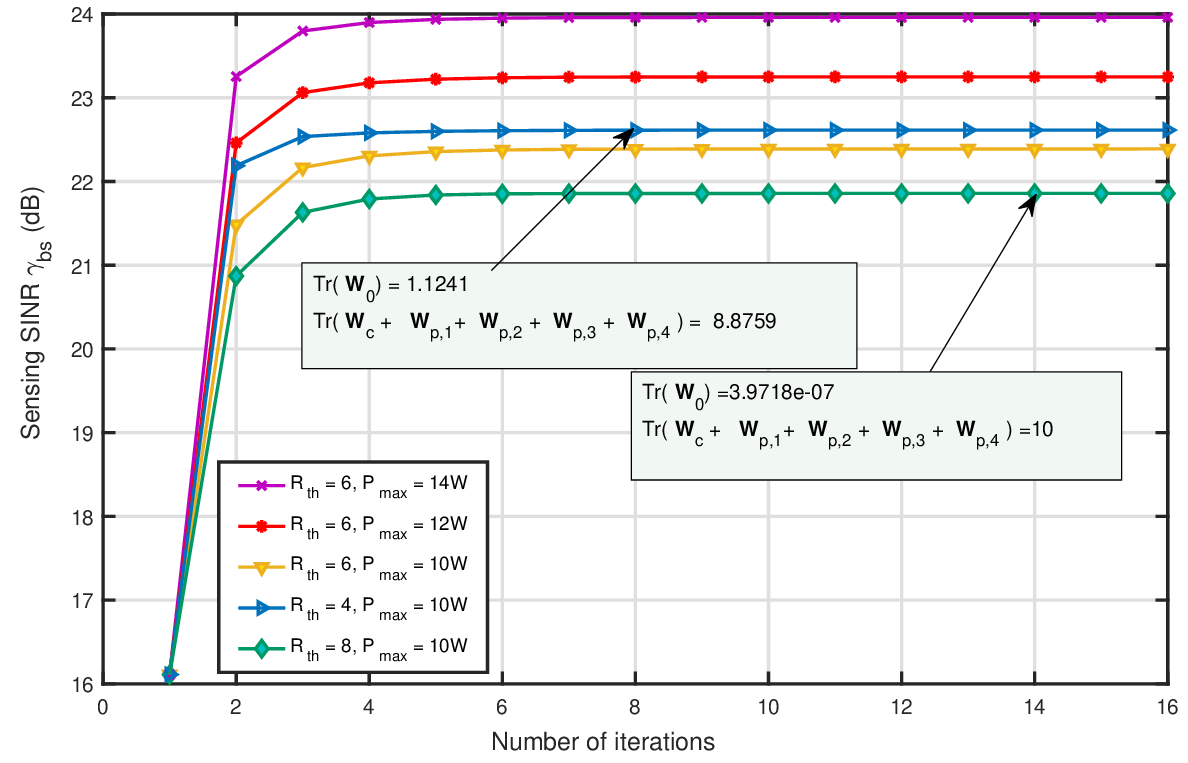}
\caption{Convergence performances under different rate thresholds and transmit powers.}
\label{SINR_versus_R}
\end{figure}

Fig.~\ref{SINR_versus_R}
depicts the achieved sensing SINR and
the convergence behaviors of the proposed overall iterative algorithm
versus various rate threshold $R_{\rm th}$ and
different available power at BS.
It shows that
the proposed algorithm converges well within few iterations.
Moreover,
it is evident that in scenarios where the BS has high available power and the GTs have a low rate threshold,
a higher sensing SINR can be achieved.
Specifically,
when $R_{\rm th} = 4$,
we have ${\rm Tr}({\bf W}_0)$ =1.1241
and when $R_{\rm th} = 8$, we have ${\rm Tr}({\bf W}_0)$ =3.9718$\times 10^{-7}$.
The reasons are explained as follows.
When the rate threshold for each GT is relatively low,
the GTs may have an increased tolerance for interference from sensing signals,
and more power at the BS is used for forming sensing signals.
In this case,
${\bf W}_c$, ${\bf W}_{{\rm p},k}$, and ${\boldsymbol h}_t$ align more closely with each other
to provide more sensing DoF and maximize the sensing SINR.
On the contrary,
when $R_{\rm th}$ is high,
most of the available power may be used to form information signals to satisfy higher communication rate requirements,
thereby resulting in fewer sensing signals being imposed on the sensing target
since they may introduce harmful interference to the GTs.

\begin{figure}[t]
 \centering
\includegraphics[width=0.48\textwidth]{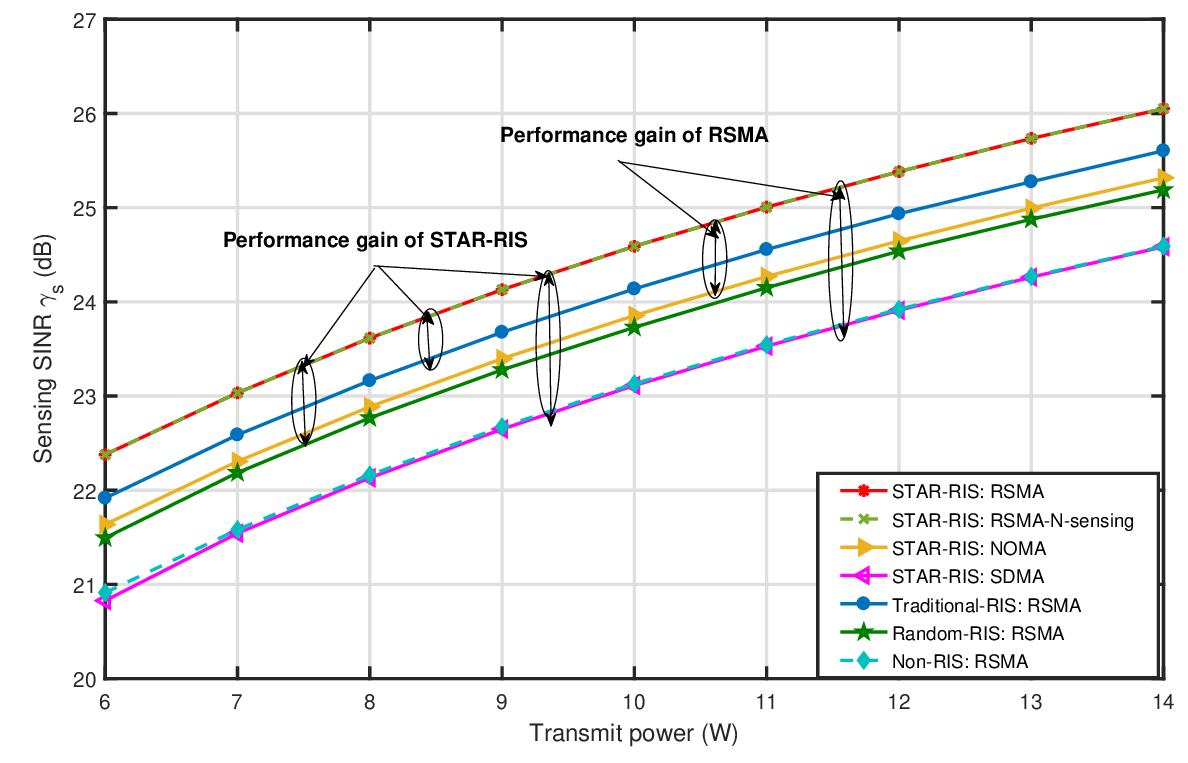}
\caption{Sensing SINR versus transmit power with different system designs.}
\label{noma_rsma_sdma_comp}
\end{figure}

Fig.~\ref{noma_rsma_sdma_comp}
compares the attained sensing SINR against the transmit power of our proposed design with other benchmark schemes,
i.e.,
the STAR-RIS aided network with RSMA scheme without sensing signals,
the STAR-RIS aided network involving NOMA/SDMA scheme,
and the traditional/Random/Non RIS aided network with RSMA scheme.
It is observable that as transmit power increases, there is a notable enhancement in the attained sensing SINR across all system designs.
Besides,
the {\bf STAR-RIS: RSMA} and {\bf STAR-RIS: RSMA-N-sensing} schemes
yield the same $\gamma_{bs}$,
which is consistent with the conclusion of Theorem \ref{thm:my1}.
Furthermore,
it is demonstrated that
irrespective of the energy consumption of STAR-RIS and traditional RISs,
the proposed STAR-RIS integrated with the RSMA approach
consumes less power than those of all other benchmarks
while achieving the same sensing SINR.
The reasons may be explained as follows.
Firstly,
with the RSMA-based scheme,
the inter-user interference can be effectively mitigated
compared with the NOMA-based scheme and SDMA-based scheme,
making it easier to meet GTs' communication rate requirements.
In this case,
the signal power allocated for sensing is stronger,
resulting in higher sensing SINR performance gain.
Secondly,
the STAR-RIS exhibits a remarkable capability to reshape beam propagation
through the dynamic optimization of both transmission and reflection beamforming,
ultimately leading to an enhancement in
sensing SINR compared to traditional RISs.
However,
only reflection or transmission phases
of the traditional RIS elements are optimized,
which leads to limited performance gain.
Thirdly,
in contrast to the {\bf random STAR-RIS} and {\bf Non-RIS} schemes,
our proposed scheme allows for more efficient signal propagation reconfiguration,
thereby yielding a substantial performance enhancement.
In summary,
both the RSMA-based transmission strategy design and the dynamic optimization of STAR-RIS transmission and reflection beamforming have made tremendous contributions to the performance improvement of the ISAC system.

\begin{figure}[t]
\centering
\includegraphics[width=0.5\textwidth]{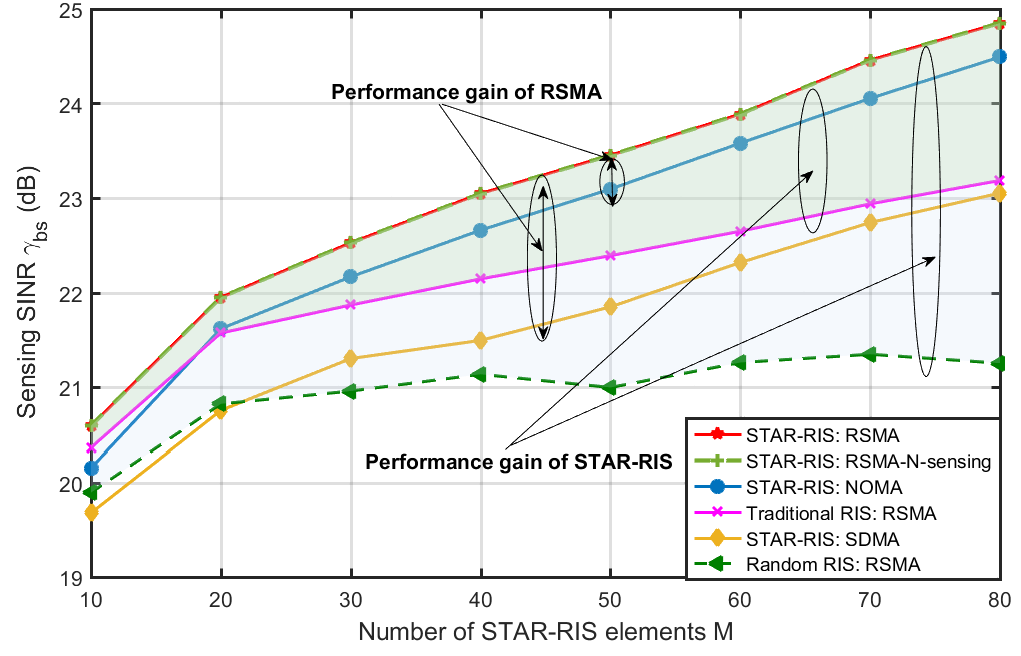}
\caption{Sensing SINR versus the number of STAR-RIS elements $M$ with different system designs.}
\label{SINR_versus_M_all_schemes}
\end{figure}

Fig.~\ref{SINR_versus_M_all_schemes}
presents the achieved maximal sensing SINR concerning the number of STAR-RIS elements $M$ across various system designs.
It is observable that as the number of STAR-RIS elements increases,
the system's sensing performance progressively enhances.
Similarly,
the results indicate that
the sensing SINR performance gain achieved through
the combination of STAR-RIS and RSMA scheme
surpasses that of other benchmarks,
aligning with the conclusions in Fig.~\ref{noma_rsma_sdma_comp}.
The range of sensing performance gains that are fulfilled through the application of STAR-RIS and RSMA schemes are also illustrated in this figure.
Notably,
with an increase in the number of STAR-RIS elements,
the performance gain gap between the {\bf STAR-RIS} scheme and the {\bf Traditional RIS }/ {\bf Random RIS} schemes progressively widens.
The main factor driving this performance enhancement is that
the STAR-RIS can manipulate both transmission and reflection beamforming simultaneously.
Conversely,
the sensing performance gains achieved by traditional RISs
are hindered by their reliance on phase shift optimization alone.
It is worth mentioning that
in the {\bf Random RIS} scheme,
the TCs and RCs of STAR-RIS are generated randomly.
In some cases,
the randomly generated TCs and RCs may align closely with the sensing channel ${\bf h}_t$, which can positively impact system performance.
However,
due to the randomness of parameter generation and the alignments,
there exists a certain degree of performance fluctuation.
In addition,
When the ISAC network transmits only communication waveforms,
it achieves the same sensing SINR as when transmitting a combination of communication and dedicated sensing waveforms.
This observation confirms the conclusions stated in Theorem \ref{thm:my1}.

\begin{figure}[t]
\centering
\includegraphics[width=0.5\textwidth]{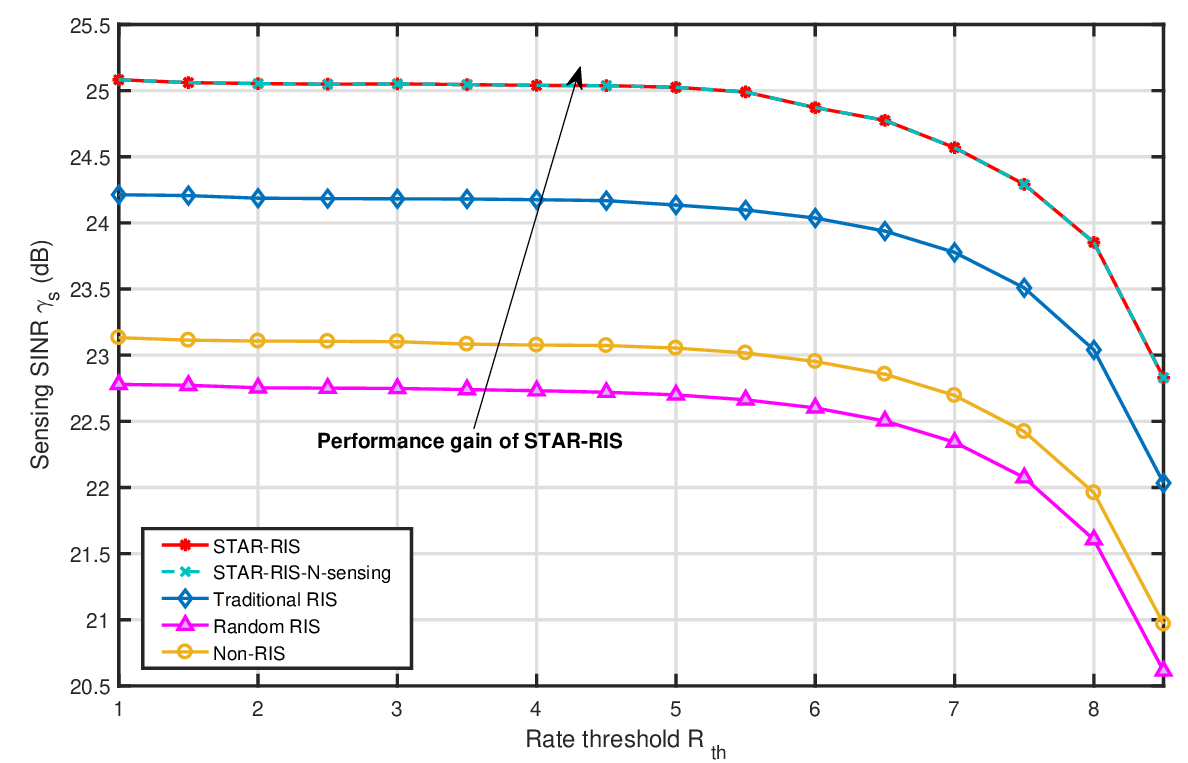}
\caption{Relationship between sensing SINR and rate threshold across various system designs.}
\label{SINR_versus_R_all_schemes}
\end{figure}

Fig.~\ref{SINR_versus_R_all_schemes}
characterizes the relationship between the
optimized sensing SINR and communication rate threshold $R_{\rm th}$ with various system designs.
One can see that as the rate threshold rises,
the sensing SINR attained by all system designs experiences a decline.
Intuitively, as the rate threshold gradually increases,
a greater amount of energy at the BS is utilized to meet the heightened communication demands,
ultimately leading to a deterioration in the achievable sensing SINR.
Besides,
the integration of STAR-RIS into networks results in a remarkable enhancement in sensing SINR, surpassing that achieved by other schemes.
Consistent with the finding presented in Theorem \ref{thm:my1} and results in Fig.~\ref{noma_rsma_sdma_comp},
irrespective of $R_{\rm th}$,
the ISAC network consistently maintains the same sensing SINR,
whether or not dedicated sensing signals are incorporated.

\begin{figure}[t]
 \centering
\includegraphics[width=0.49\textwidth]{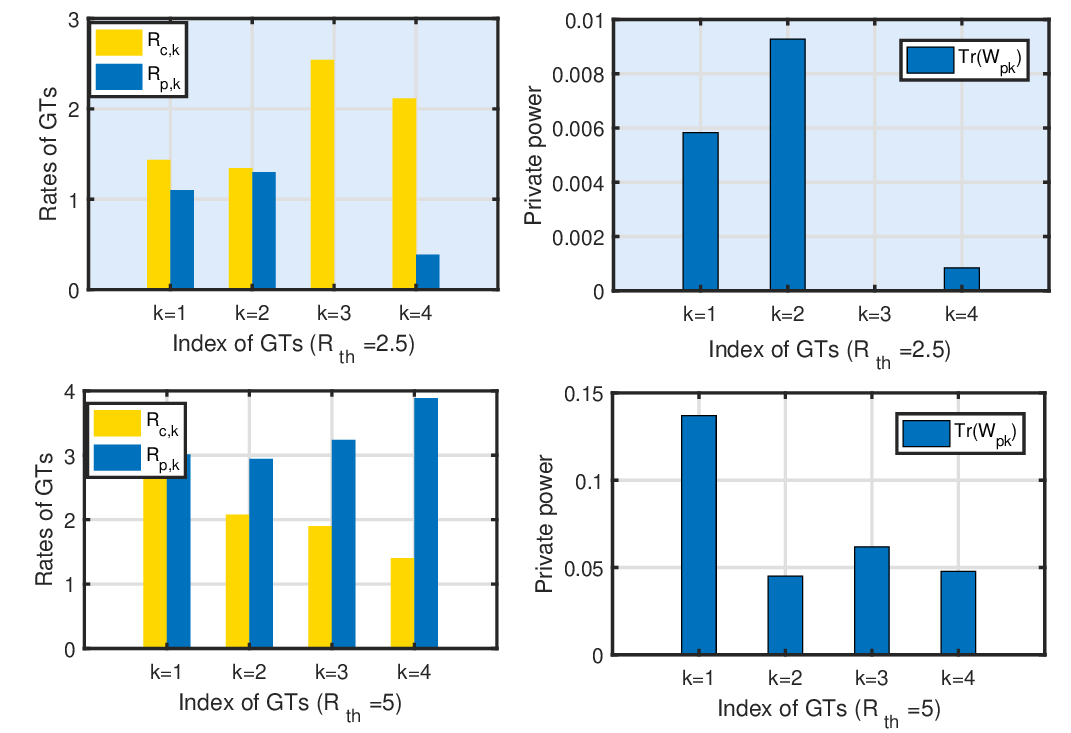}
\caption{Optimized common rates, private rates and private power of each GT.}
\label{bar_two_GT_rates_power_show}
\end{figure}

Fig.~\ref{bar_two_GT_rates_power_show}
provides the common rates and private rates of four GTs,
and the corresponding private power allocation of GTs.
It shows that the optimized results of private rates are consistent with the corresponding optimized power allocation.
In other words,
when the allocated common stream can meet the rate requirement of the GT, no private information stream will be allocated to it, and the corresponding private rate is 0.
Furthermore,
when the rate requirement of each GT is more stringent,
a greater amount of power will be assigned to the private information stream to fulfill the rate requirement,
thereby leading to the private rates of all GTs increasing.

% µ±ËÙÂÊãÐÖµ±È½ÏÐ¡Ê±£¬ÓÃ»§¶ÔÓÚ¸ÉÈÅµÄÈÝÈÌ³Ì¶È½Ï¸ß£¬Í¨¹ýÀûÓÃ¹«ÓÐÐÅÏ¢Á÷¾Í×ãÒÔÂú×ãÓÃ»§µÄÍ¨ÐÅÒªÇó¡£µ±ËÙÂÊãÐÖµÔö´óµÄÊ±ºò£¬ËäÈ»Í¨¹ýÀûÓÃ¹«ÓÐÐÅÏ¢¿ÉÒÔ½µµÍÒ»²¿·Ö¸ÉÈÅ£¬µ«ÊÇÆä¹«ÓÐÐÅÏ¢Á÷¶ÔÓÚÍ¨ÐÅËÙÂÊÐÔÄÜµÄ¹±Ï×ÊÇÓÐÏÞµÄ£¬±ØÐëÔÚÊ¹ÓÃ¹«ÓÐÐÅÏ¢Á÷½µµÍÒ»¶¨³Ì¶ÈµÄ¸ÉÈÅºó£¬Í¨¹ýÌáÉýË½ÓÐËÙÂÊ²ÅÄÜÂú×ã¸÷¸öÓÃ»§µÄËÙÂÊÒªÇó¡£
% »»Ò»ÏÂÓÃ»§Î»ÖÃÊÔÊÔ¿´£¬ÅÜÒ»×éÊý¾Ý¿´¿´

\begin{figure}[t]
 \centering
\includegraphics[width=0.48\textwidth]{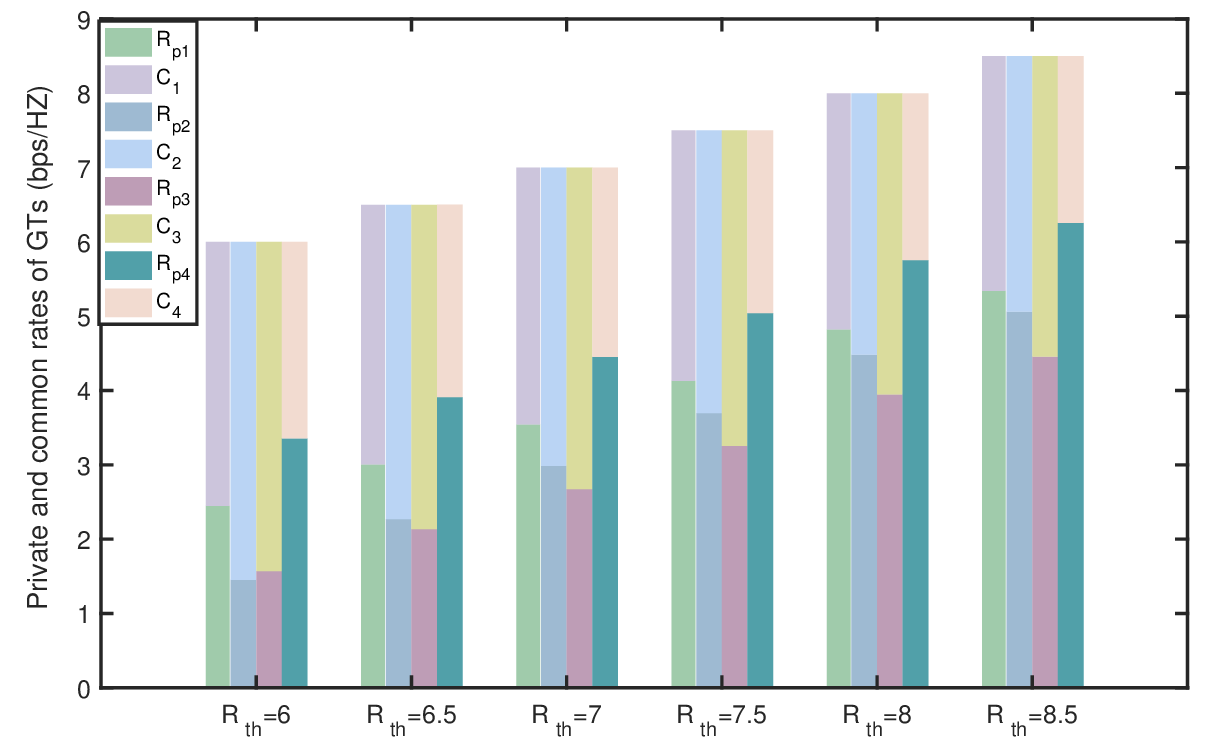}
\caption{Common rates and private rates of GTs versus the rate thresholds.}
\label{rate_com_private_bar_figure}
\end{figure}

Fig.~\ref{rate_com_private_bar_figure}
depicts the common rates and private rates of four GTs versus various
rate thresholds.
It indicates that with an increase in the rate threshold,
the proportion of private rate for each GT within the total rate also rises.
However,
the variation in common rates for each GT is not particularly evident as the rate threshold increases.
The reason could be that,
at a relatively low rate threshold,
each GT exhibits a higher tolerance for inter-interference,
and the communication rate requirements of each GT
can be relatively easily satisfied through the utilization of the common information flow.
Nonetheless,
when the communication rate threshold is high,
despite the capability to attain flexible interference management by dividing user messages into common and corresponding private messages,
the effect of allocating common information on interference suppression and rate enhancement is limited.
On the other words,
after mitigating interference to a certain extent by leveraging the common information flow,
it becomes crucial to increase private rates to satisfy the specific rate requirements of each GT.

\begin{figure}[t]
\centering
\subfigcapskip=-10pt %ÉèÖÃ×ÓÍ¼Óë×Ó±êÌâÖ®¼äµÄ¾àÀë
\subfigure[$R_{\rm th}$ = 6]{
\begin{minipage}[t]{0.98\linewidth}
\centering
\includegraphics[width=1\textwidth]{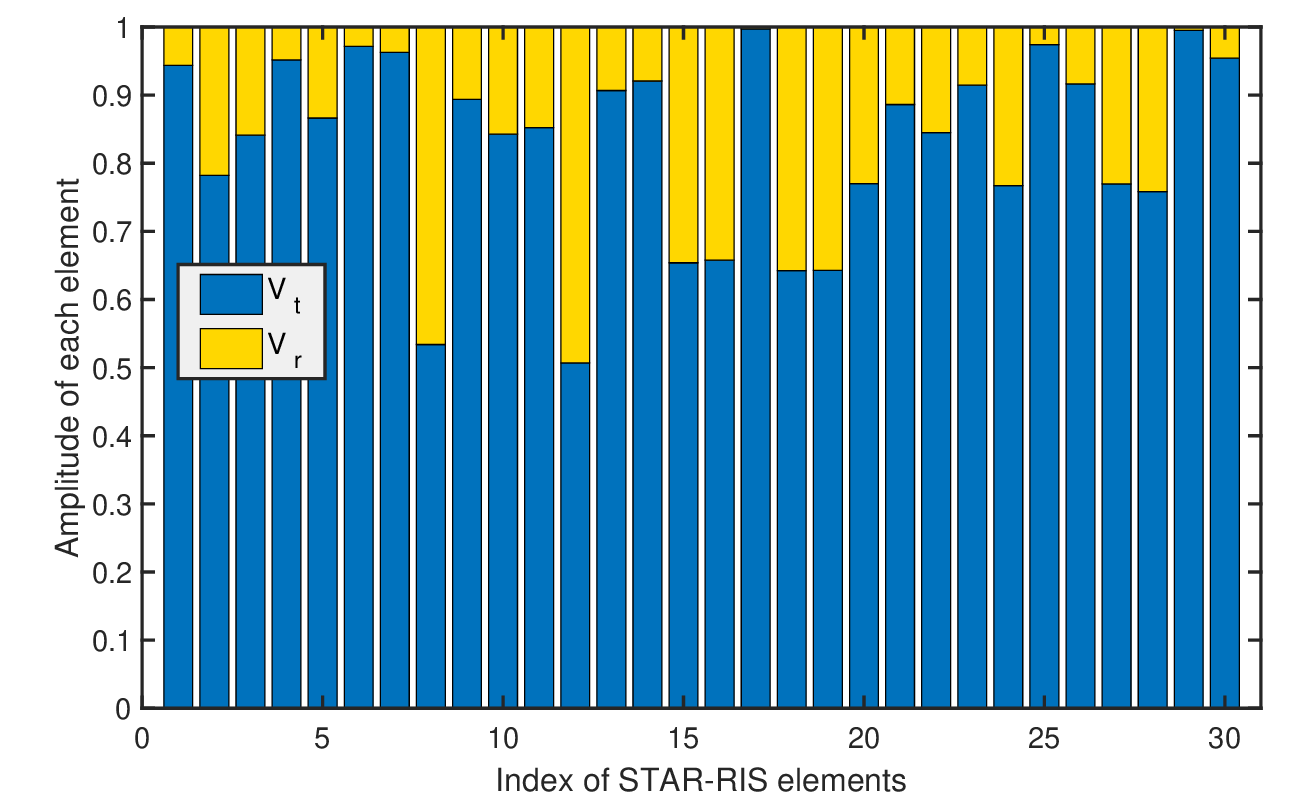}
\label{amplitude_R_6}
\end{minipage}
}
\\
\subfigure[$R_{\rm th}$ = 1]{
\begin{minipage}[t]{0.98\linewidth}
\centering
\includegraphics[width=1\textwidth]{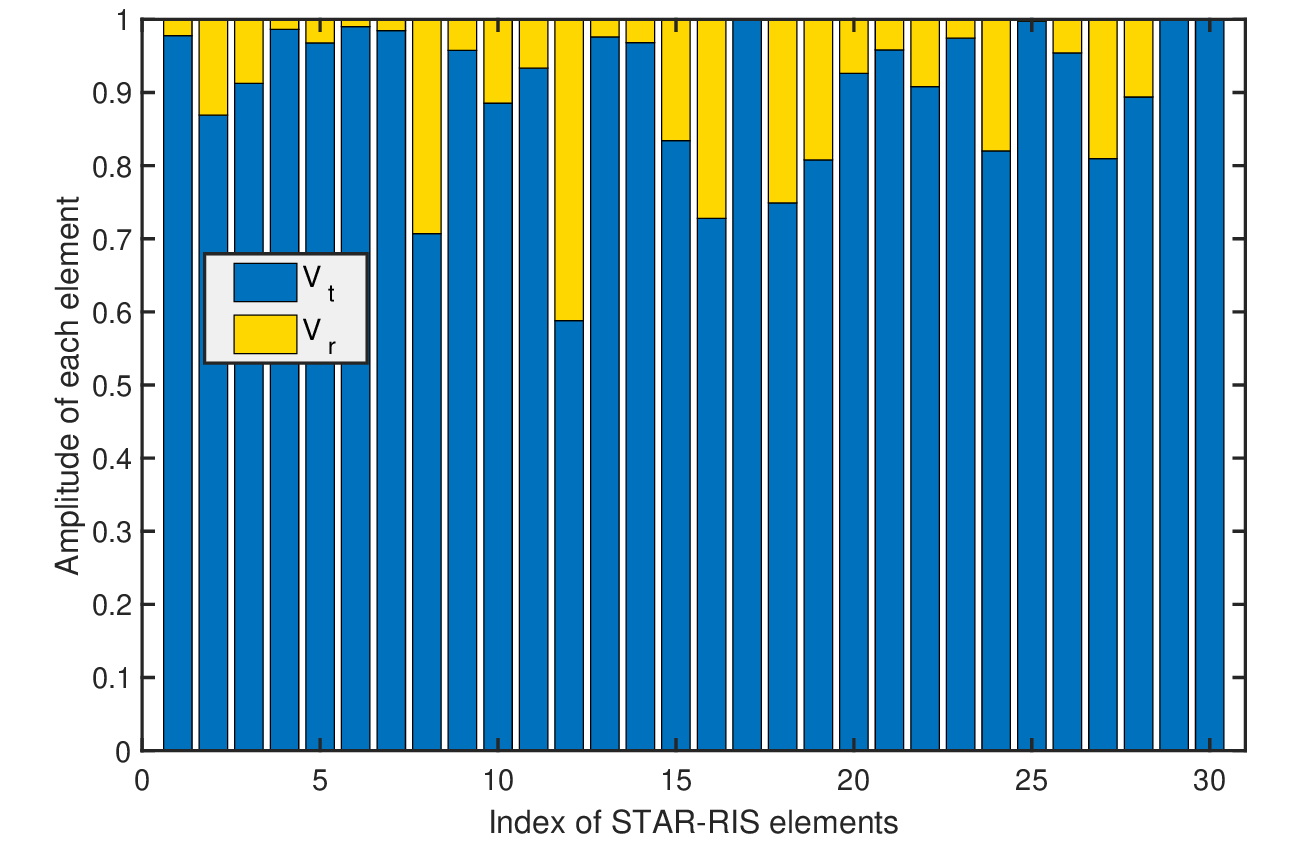}
\label{amplitude_R_1}
\end{minipage}
}
\caption{
Optimized amplitudes for every STAR-RIS element across different rate threshold.}
\label{amplitude_of_IRS}
\end{figure}

Fig.~\ref{amplitude_of_IRS}
provides the optimized amplitudes for transmission and reflection of every STAR-RIS element across various rate thresholds $R_{\rm th}$.
It demonstrates that every element possesses the capability to adaptively manipulate the incoming signals based on the corresponding channel condition.
In addition,
the transmission amplitude coefficients are relatively larger than
the reflection amplitude coefficients,
which means that more energy is shared for enhancing the sensing performance.
Besides,
comparing the optimized results shown in Fig.~\ref{amplitude_R_6}
with that in Fig.~\ref{amplitude_R_1},
it is noticed that
the reflection amplitude coefficients of each STAR-RIS element
tend to be relatively greater when the rate threshold $R_{\rm th}$ is high.
The reason is
that when the rate requirements of GTs are large,
the BS has to distribute a greater amount of energy into reflection space,
so as to achieve signal enhancement for each GT.

%\begin{figure}[!t]
%\centering
%\subfloat[$R_{\rm th}$ = 6]{\includegraphics[width=2.5in]{figs/amplitude_of_each_element_R_6.eps}%
%\label{amplitude_R_6}}
%% \hfil
%\\
%\subfloat[$R_{\rm th}$ = 1]{\includegraphics[width=2.5in]{figs/amplitude_of_each_element_R_1.eps}%
%\label{amplitude_R_1}}
%\caption{
%Optimized amplitudes for every STAR-RIS element across different rate threshold.}
%\label{amplitude_of_IRS}
%\end{figure}
%
%
%%Fig.~\ref{beam_pattern_versus_P_R}
%%characterizes the optimized beampattern with different rate threshold and
%%available power,
%%where the sensing target is located at the angle direction of $10^{\circ}$.
%%It is observed that the main beam consistently aligns with the direction of the intended target across various parameter settings,
%%thereby validating the effectiveness of the proposed beamforming design for target detection.
%%Besides,
%%when the available power at the BS is substantial,
%%the optimized beampattern gain is higher, resulting in stronger signal strength.
%%In addition,
%%the beampattern with a higher peak-to-sidelobe ratio
%%can be achieved when $R_{\rm th} = 4$,
%%which indicates that the signal energy is predominantly concentrated in the main beam direction,
%%aiming at improving the accuracy of target detection.
%%\begin{figure}[t]
%% \centering
%%\includegraphics[width=0.48\textwidth]{figs/fig9_beam_pattern_verus_R_P.eps}
%%\caption{Transmit-receive beampattern with different rate threshold and available power .}
%%\label{beam_pattern_versus_P_R}
%%\end{figure}

\section{Conclusion}\label{conclusion}
In this paper,
we have explored
the integration of STAR-RIS into an ISAC system employing the RSMA-based approach,
where the maximization of sensing SINR was
attained through the unified design of rate
splitting in conjunction with precise adjustments of beamforming
at BS and STAR-RIS, respectively.
To address the formulated non-convex problem, the SCA-based iterative algorithm incorporating SDR, MM, and SROCR techniques has been devised.
Simulation results manifest that,
in comparison to other benchmark schemes,
the employment of the RSMA scheme
in conjunction with the transmission and reflection beamforming design at the STAR-RIS is crucial for enhancing the system's sensing performance.
Additionally, there is a balance to be established between achieving high-sensing SINR and maintaining reliable communication performance.
For the considered ISAC network with a single target,
the achievable sensing SINR is the same regardless of whether specially designed sensing signals are incorporated or not.
For the mobile scenario involving multiple sensing targets and users in the STAR-RIS-enhanced ISAC system, 
we will investigate efficient sensing methods,
specifically Time division (TD) sensing, signature sequence (SS) sensing, and hybrid TD-SS sensing.

\begin{appendices}
\section{{Proof of Theorem \ref{thm:my1}}}\label{App:T1}
According to the optimized solution
$\{\widehat{{\bf W}}_c, \widehat{{\bf W}}_{{\rm p},k}, \widehat{{\bf W}}_0\}$ for
sub-problem ${\mathbf {P}}_{\mathbf {B_1}}$,
we can reconstruct one set of new solution $\{\overline{{\bf W}}_c,\overline{{\bf W}}_{{\rm p},k}, \overline{{\bf W}}_{0}\}$,
which are mathematical modeled as
\begin{equation}
\begin{aligned}
&\overline{{\bf W}}_c = \widehat{{\bf W}}_c + \zeta_c \widehat{{\bf W}}_0,~
\overline{{\bf W}}_{{\rm p},k} = \widehat{{\bf W}}_{{\rm p},k} + \zeta_k \widehat{{\bf W}}_0,\\
&\overline{{\bf W}}_{0} = 0,~~\zeta_c + \sum \nolimits_{k=1}^K \zeta_k = 1.
\label{eq_new_solutions}
\end{aligned}
\end{equation}
By substituting the new set of solution into constraint (\ref{cs_2}),
we have
\begin{align*}
&R_{c,k} = \log_2\left(1+\frac{{\rm Tr}( {\bf H}_k \overline{{\bf W}}_c )}{
\sum \nolimits_{i=1}^K {\rm Tr}({\bf H}_k \overline{{\bf W}}_{{\rm p},i})  +\sigma_k^2}\right) \nonumber \\
%&=\log_2\left(1+\frac{{\rm Tr}( {\bf H}_k (\widehat{{\bf W}}_c+ \zeta_c \widehat{{\bf W}}_0)) }{
%\sum \nolimits_{i=1}^K {\rm Tr}({\bf H}_k (\widehat{{\bf W}}_{p,i} + \zeta_i \widehat{{\bf W}}_0))  +n_k}\right) \nonumber \\
& = \log_2\left(
\tfrac{{\rm Tr}( {\bf H}_k (
\widehat{{\bf W}}_c + \zeta_c \widehat{{\bf W}}_0 +
\sum \limits_{i=1}^K (\widehat{{\bf W}}_{{\rm p},i} + \zeta_i \widehat{{\bf W}}_0)))
+ \sigma_k^2
}
{\sum \nolimits_{i=1}^K {\rm Tr}({\bf H}_k (\widehat{{\bf W}}_{{\rm p},i} + \zeta_i
\widehat{{\bf W}}_0))  + \sigma_k^2}\right) \nonumber\\
&=
 \log_2\left(\frac{{\rm Tr}( {\bf H}_k (\widehat{{\bf W}}_c +
\sum \limits_{i=1}^K \widehat{{\bf W}}_{{\rm p},i} +  \widehat{{\bf W}}_0))
+ \sigma_k^2
}
{\sum \nolimits_{i=1}^K {\rm Tr}
({\bf H}_k (\widehat{{\bf W}}_{{\rm p},i} + \zeta_i \widehat{{\bf W}}_0))  +\sigma_k^2}\right) \nonumber \\
&
\overset{(a)}{\ge}
\log_2\left(\frac{{\rm Tr}( {\bf H}_k (\widehat{{\bf W}}_c +
\sum \limits_{i=1}^K \widehat{{\bf W}}_{{\rm p},i} +  \widehat{{\bf W}}_0))
+ \sigma_k^2
}
{\sum \nolimits_{i=1}^K {\rm Tr}({\bf H}_k (\widehat{{\bf W}}_{{\rm p},i} +\widehat{{\bf W}}_0))  +\sigma_k^2}\right) \nonumber \\
& = \log_2\left(1+\frac{{\rm Tr}( {\bf H}_k \widehat{{\bf W}}_c )}{\sum \nolimits_{i=1}^K {\rm Tr}({\bf H}_k (\widehat{{\bf W}}_{{\rm p},i}+ \widehat{{\bf W}}_0) ) +\sigma_k^2}\right) \nonumber \\
& \overset{(b)}{\ge}
\sum \nolimits_{k=1}^K c_k.
\end{align*}
The step $(a)$ holds due to the fact that $\sum \nolimits_{i=1}^K\zeta_i \le 1$,
and step $(b)$ holds since
$\{\widehat{{\bf W}}_c, \widehat{{\bf W}}_{{\rm p},k}, \widehat{{\bf W}}_0\}$ is one set of feasible solutions for ${\mathbf {P}}_{\textbf {B1-1}}$.
On the other hand,
by substituting the new set of solutions into constraint (\ref{cs_3}),
we have
\begin{align*}
&R_{{\rm p},k} =
\log_2\left(1+
\frac{{\rm Tr}( {\bf H}_k \overline{{\bf W}}_{{\rm p},k} )}{
\sum \nolimits_{j\neq k}^K {\rm Tr}({\bf H}_k \overline{{\bf W}}_{{\rm p},j})  +\sigma_k^2}\right) \nonumber \\
%&=\log_2\left(1+
%\frac{{\rm Tr}( {\bf H}_k (\widehat{{\bf W}}_{p,k} + \zeta_k \widehat{{\bf W}}_0))}{
%\sum \nolimits_{j\neq k}^K {\rm Tr}({\bf H}_k (\widehat{{\bf W}}_{p,j} +\zeta_j \widehat{{\bf W}}_0) )  + n_k}\right) \nonumber \\
&=\log_2\left(
\frac{
{\rm Tr}({\bf H}_k (\sum \nolimits_{i=1}^K \widehat{{\bf W}}_{{\rm p},i} + \widehat{{\bf W}}_0))}
{
\sum \nolimits_{j\neq k}^K
{\rm Tr}({\bf H}_k (\widehat{{\bf W}}_{{\rm p},j} +\zeta_j \widehat{{\bf W}}_0) )  + \sigma_k^2}\right) \nonumber \\
&\overset{(c)}{\ge}
\log_2\left(
\frac{
{\rm Tr}({\bf H}_k (\sum \nolimits_{i=1}^K \widehat{{\bf W}}_{{\rm p},i} + \widehat{{\bf W}}_0))}
{\sum \nolimits_{j\neq k}^K {\rm Tr}({\bf H}_k (\widehat{{\bf W}}_{{\rm p},j} + \widehat{{\bf W}}_0) )  + \sigma_k^2}\right) \nonumber \\
& =\log_2\left(1+
\frac{{\rm Tr}( {\bf H}_k \widehat{{\bf W}}_{{\rm p},k} )}{
\sum \nolimits_{j\neq k}^K {\rm Tr}(
{\bf H}_k (\widehat{{\bf W}}_{{\rm p},j}  + \widehat{{\bf W}}_0)) + \sigma_k^2}\right) \nonumber \\
& \overset{(d)}{\ge} R_{k}^{\rm th} -c_k.
\end{align*}
The validity of step $(c)$ is attributed to the fact that
$\sum \nolimits_{j\neq k}^K\zeta_j \le 1$,
and step $(d)$ holds since
$\{\widehat{{\bf W}}_c, \widehat{{\bf W}}_{{\rm p},k}, \widehat{{\bf W}}_0\}$ is one set of feasible solutions for ${\mathbf {P}}_{\textbf {B1-1}}$.
Therefore,
$\{\overline{{\bf W}}_c,\overline{{\bf W}}_{{\rm p},k}, \overline{{\bf W}}_{0}\}$ is also one set
of feasible solutions for ${\mathbf {P}}_{\textbf {B1-1}}$.
It can also achieve the optimal sensing SINR,
which is not inferior to that achieved by $\{\widehat{{\bf W}}_c, \widehat{{\bf W}}_{{\rm p},k}, \widehat{{\bf W}}_0\}$.
The proof for Theorem \ref{thm:my1} is completed.
\end{appendices}

% \vfill

\end{document}